Fachhochschule der Wirtschaft

-FHDW-

Bergisch Gladbach

Diplomarbeit

Thema:

Verbesserung von OS- und Service-Fingerprinting mittels Fuzzing

Prüfer:

Prof. Dr. Ralf Schumann

Dr. Thomas Seifert

Verfasser:

Michael Hanspach

Studiengang Wirtschaftsinformatik

Eingereicht am:

22. September 2008

# Executive Summary

Im Rahmen eines nachhaltigen Risikomanagements sind Unternehmen gezwungen, ihre IT-Infrastruktur vor sicherheitskritischen Schwachstellen zu schützen. Im Auftrag des Kunden werden simulierte Einbruchsversuche (Penetrations-Tests) in IT-Systemen durchgeführt, um eine solide Grundlage für die Bewertung des erreichten Sicherheitsniveaus zu schaffen und dem Kunden eine umfassende Beratung bei der Behebung von Schwachstellen zu bieten.

Für die Durchführung eines Penetration-Tests ist zunächst die Identifizierung von Betriebssystemen, Protokollen und Diensten auf den Zielsystemen, das sogenannte Fingerprinting, von zentraler Bedeutung. Ziel der Diplomarbeit ist es, mögliche Verbesserungspotentiale des Fingerprintings zu erforschen. Dies soll mit dem Einsatz von Fuzzing, der systematischen Anfertigung von Anfrage-Nachrichten, erreicht werden. Durch Versenden der so präparierten Anfragen an unterschiedliche Hosts sollen subtile Unterschiede in den Antwort-Nachrichten erfasst werden, die ein Identifizierungsmerkmal bestimmter Betriebssysteme und Dienste bilden.

Auf dieser Grundlage entwickelt der Verfasser ein eigenes Konzept für ein Fingerprinting durch Anwendung von Fuzzing. Dabei zerlegt er die Gesamtaufgabe zunächst in Teilschritte und fügt diese in ein Vorgehensmodell ein. Anschließend beschreibt und begründet er für jeden Schritt das gewählte Vorgehen, die eingesetzten Methoden und die entworfenen Algorithmen. Zum Beweis der Durchführbarkeit werden zwei Fallstudien mit einer eigens entwickelten Fingerprinting-Software auf Basis bestehender Programmbibliotheken durchgeführt. Im Kontext der Fallstudien wird die geschaffene Lösung getestet, bewertet und mit bestehenden Ansätzen verglichen.

Im Ergebnis der Diplomarbeit wurden neue Unterscheidungsmerkmale für Betriebssysteme und Dienste erfasst, die sich zur Verbesserung bestehender und zukünftiger Fingerprinting-Software eignen.

# Inhaltsverzeichnis

# Abbildungsverzeichnis

# Tabellenverzeichnis

# Abkürzungsverzeichnis

| | |
|---|---|
| API | Application Programming Interface |
| BDSG | Bundesdatenschutzgesetz |
| BSI | Bundesamt für Sicherheit in der Informationstechnik |
| DARPA | Defense Advanced Research Project Agency |
| DNS | Domain Name System |
| FTP | File Transfer Protocol |
| GUI | Graphical User Interface |
| HTTP | Hypertext Transfer Protocol |
| IANA | Internet Assigned Numbers Authority |
| IETF | Internet Engineering Task Force |
| IP | Internet Protocol |
| ISN | Initial Sequence Number |
| KonTraG | Gesetz zur Kontrolle und Transparenz im Unternehmensbereich |
| LAN | Local Area Network |
| LOC | Lines of Code |
| NAT | Network Address Translation |
| OS | Operating System |
| PDU | Protocol Data Unit |
| RAD | Rapid Application Development |
| RFC | Request for Comment |
| TCP | Transmission Control Protocol |
| UML | Unified Modeling Language |
| URI | Uniform Resource Identificator |
| URL | Uniform Resource Locator |
| WAN | Wide Area Network |

# 1 Einleitung

## *1.1 Motivation*

Die Informationstechnik hat die Gesellschaft im Allgemeinen und die Unternehmenslandschaft im Besonderen nachhaltig verändert. Kaum ein Unternehmen ist noch ohne den Einsatz elektronischer Datenverarbeitung denkbar. Der Begriff des Ubiquitous Computing, der allgegenwärtigen rechnergestützten Informationsverarbeitung, beschreibt die Rolle der Informationstechnik im 21. Jahrhundert besonders treffend.

Im Zuge dieser Entwicklung steigen die Ansprüche an das IT-Sicherheitsniveau von Unternehmen stetig. Es gilt zuverlässig zu verhindern, dass die eingesetzte IT-Infrastruktur durch menschliche (fahrlässiges oder vorsätzliches Handeln von Personen) oder natürliche Faktoren (Naturereignisse) beeinträchtigt wird. So kann ein ausgefallener Server ohne angemessenes Risikomanagement und Notfallvorsorge vorübergehend oder dauerhaft die gesamte Geschäftstätigkeit stilllegen. Auch Wirtschaftsspionage und der Verlust oder die Manipulation von kritischen Daten sind Szenarien, die hohe Kosten nach sich ziehen können. Zudem können gravierende Sicherheitsprobleme für ein Unternehmen zu erheblichen Imageschäden in der Öffentlichkeit führen und bergen ein gewisses Risiko, in Bezug auf die Kreditwürdigkeit herabgestuft zu werden. Doch auch abseits der Wahrnehmung dieser natürlichen Interessen gibt es genügend gesetzliche Vorschriften, die die Umsetzung von IT-Sicherheitsmaßnahmen in Unternehmen einfordern.[1]

So kann die Geschäftsleitung von Kapitalgesellschaften[2] durchaus mit ihrem Privatvermögen für Schäden, die durch mangelnde Umsetzung von IT-Sicherheitsmaßnahmen entstanden sind, zur Verantwortung gezogen werden.[3] Auch fehlende Spezialkenntnisse im Bereich IT-Sicherheit haben keinen Einfluss auf die Haftung der Verantwortlichen.[4] Weitere zivilrechtliche Schadensersatzansprüche gegen Unternehmen ergeben sich insbesondere aus dem Bundesdatenschutzgesetz (BDSG) und dem Gesetz zur Kontrolle und Transparenz im Unternehmensbereich (KonTraG), in dem der Vorstand von Kapitalgesellschaften zur Schaffung eines Risikomanagementsystems verpflichtet wird, durch das mögliche Sicherheitsschwachstellen schon im Vorfeld erkannt werden sollen.[5]

---

[1]    Vgl. Mörike, Michael (Hrsg.) (2004), S. 76
[2]    Gegenüber Kapitalgesellschaften ist die private Haftung von Kaufleuten und Gesellschaftern von Personengesellschaften schon aufgrund ihrer Rechtsform gegeben.
[3]    Vgl. §§ 43 Abs1. GmbHG und 93 Abs. 1 AktG
[4]    Vgl. Speichert, Horst (2007), S. 225f

Oben genannte Anspruchsgrundlagen werfen die Frage auf, wie die Umsetzung geeigneter IT-Sicherheitsmaßnahmen erfolgreich nachgewiesen werden kann, um rechtliche Konsequenzen zu vermeiden. Zu diesem Zweck werden Penetrations-Tests durch externe Dienstleister durchgeführt, in denen die Denk- und Vorgehensweise eines echten Angreifers übernommen werden soll, der versucht, die Sicherheit des Unternehmens zu gefährden.[6]

## *1.2 Ziel der Diplomarbeit*

Im Rahmen von Penetrations-Tests gilt es, die Sicherheitsmaßnahmen eines Unternehmens zu durchbrechen. Zu diesem Zweck existieren sowohl technische als auch psychologische Mittel (Social Engineering). Die vorliegende Diplomarbeit beschränkt sich jedoch auf die technischen Möglichkeiten, in ein System einzudringen. Angriffe auf Zielsysteme erfolgen dabei über das Netzwerk, zum einen, weil die meisten Zielsysteme nicht physisch zugänglich (etwa in einem Serverraum verschlossen) sind, vor allem aber auch, weil dies eines der gefürchtetsten Angriffsszenarien ist.[7]

Bei diesem Vorhaben spielt die Analyse der entfernten Rechner eine wichtige Rolle. Die Unterscheidung zwischen verschiedenen Diensten und Betriebssystemen (Fingerprinting) spielt nicht nur in Bezug auf die bekannten Schwachstellen, sondern auch bei der Wahl der grundlegenden Angriffsweisen eine große Rolle.[8]

Für den Einsatz in Penetrations-Tests sind die bestehenden Fingerprinting-Tools zwar nützlich, jedoch bislang nicht genau genug. Schwächen gibt es vor allem bei der Differenzierung zwischen unterschiedlichen Releases des gleichen Betriebssystem bzw. des gleichen Dienstes. Die Genauigkeit des Fingerprintings kann durch eine größere Zahl von Testfällen gegenüber bestehender Fingerprinting-Software verbessert werden.[9]

---

[5]   Vgl. § 91, Abs. 2 AktG
[6]   Vgl. Mörike, Michael (Hrsg.) (2004), S. 76
[7]   Vgl. Johansson und Riley (2005)
[8]   Vgl. BSI (2003)

Diese Erkenntnisse werden auch durch die in Kapitel 5 durchgeführten Tests bestätigt. So gelingt es dem Verfasser im Rahmen der Diplomarbeit, neue Testfälle zu finden, mit denen sich die Genauigkeit von Fingerprinting-Software verbessern lässt.

Dies geschieht mittels der Anwendung von *Fuzzing* – hier: dem Versenden von systematisch erzeugten Nachrichten. So können bestimmte Verhaltensweisen von Zielsystemen provoziert werden, die durch die manuelle Erzeugung von Nachrichten nicht entdeckt worden wären.

Ziel der Diplomarbeit war es, neue Anfrage-Antwort-Kombinationen zu finden, die sich in verschiedenen Betriebssystemen und Diensten voneinander unterscheiden. Diese Kombinationen bieten so wie menschliche Fingerabdrücke ein Identifizierungsmerkmal.

Die Verknüpfung von Fingerprinting und Fuzzing ist nach bester Kenntnis des Verfassers eine Neuschöpfung. In diesem Sinne behandelt die vorliegende Diplomarbeit ein Forschungsthema mit direktem Bezug zur Arbeit von IT-Sicherheitsdienstleistern und insbesondere Penetrations-Testern.

---

[9]  Vgl. Poppa, Ryan (2007)

## *1.4 Vorgehensweise und Aufbau der Diplomarbeit*

Die vorliegende Diplomarbeit ist in 6 Kapitel gegliedert.

Kapitel 2 beinhaltet eine intensive Literaturrecherche zum Zweck, bestehende Methoden und Konzepte darzustellen, deren Verständnis für die Erarbeitung einer Lösung erforderlich ist. Zunächst werden Grundbegriffe der IT-Sicherheit und von TCP/IP-Netzwerken beschrieben. Eine Definition von Fingerprinting sowie die Analyse bekannter Fingerprinting-Ansätze und Methoden schließen sich an. Zuletzt erfolgt eine Analyse der Technik des Fuzzing, bei der verschiedene Fuzzing-Methoden erklärt und verglichen werden, um diese später zum Zweck des Fingerprintings einzusetzen.

Kapitel 3 beinhaltet das Konzept eines verbesserten Fingerprintings durch Einsatz von Fuzzing. Es kann als persönlicher Beitrag des Verfassers zur Theorie eines verbesserten Fingerprintings betrachtet werden. Das Konzept enthält die Auswahl der geeigneten Methoden des Fingerprintings und des Fuzzings, den Entwurf eines Phasenmodells für ein verbessertes Fingerprinting, die Entwicklung geeigneter Vergleichsmechanismen für ähnliche Betriebssysteme und Dienste sowie die fachlichen und technischen Anforderungen an eine Implementierung.

Kapitel 4 präsentiert die prototypische Implementierung der konzipierten Fingerprinting-Lösung. Es werden die programmiertechnischen Voraussetzungen und eingesetzten Technologien genannt sowie die konkreten Programm- und Dateistrukturen erklärt.

Kapitel 5 hat den Zweck, die Zuverlässigkeit der entwickelten Fingerprinting-Software zu testen und die Durchführbarkeit des Konzeptes im Sinne eines Proof of Concept zu belegen.

Kapitel 6 bietet schließlich eine Zusammenfassung der Vorgehensweise und Ergebnisse der Diplomarbeit im Rückblick. Der Verfasser zieht ein persönliches Fazit und wagt einen Ausblick auf mögliche zukünftige Entwicklungen.

# 2 Grundlagen

## *2.1 IT-Sicherheit*

### 2.1.1 Definition und Schutzziele

Das Bundesamt für Sicherheit in der Informationstechnik (BSI) definiert IT-Sicherheit als „Schutz von Daten hinsichtlich gegebener Anforderungen an deren Vertraulichkeit, Verfügbarkeit und Integrität".[10] Jene erwähnten Anforderungen werden im Folgenden als Schutzziele bezeichnet und näher beschrieben.

- Vertraulichkeit
  Die Daten sind vor unbefugtem Zugriff geschützt.
- Verfügbarkeit
  Daten müssen zum Zeitpunkt der Anforderung zugänglich sein.
- Integrität
  Daten sollen vollständig und unverändert sein.

Dies sind nicht die einzigen denkbaren Schutzziele. Denkbar wären etwa die Schutzziele Anonymität, Nichtanfechtbarkeit und Verhinderung von Verzögerungen.[11]

IT-Sicherheitsdienstleister führen Penetrations-Tests durch, um Schwachstellen zu finden, die die Einhaltung dieser Schutzziele gefährden können.

### 2.1.2 Durchführung von Penetrations-Tests

Ein Penetrations-Test ist „die Prüfung von Systemen (Hardware, Software, Prozesse) auf Sicherheitsschwachstellen, mit den Methoden und Motivationen, die ein Angreifer (Cracker) anwenden würde, um Kontrolle über das System und seine Daten zu erlangen".[12]

---

[10] Vgl. BSI (2007), S. 8
[11] Vgl. Haake et al. (2007), S. 54f
[12] Ruef, Marc (2007), S. 15

Das BSI empfiehlt dabei folgende Vorgehensweise[13] für die Durchführung eines Penetrations-Tests.

1. Recherche nach Informationen über das Zielsystem
   Über die WHOIS-Datenbank kann der Eigentümer der IP-Adresse des Zielsystems ermittelt werden. Dadurch können möglicherweise Rückschlüsse auf die Beschaffenheit des Zielsystems gezogen werden.

2. Scan der Zielsysteme auf angebotene Dienste
   Die Zielsysteme werden einem Port-Scan (vgl. Kapitel 2.2.3) unterzogen. Evtl. geöffnete Ports ermöglichen erste Vermutungen über die dahinter stehenden Dienste.

3. System- und Anwendungserkennung
   Durch Methoden des OS- und Service-Fingerprintings wird versucht, das Betriebssystem und die spezifischen Dienste des Zielsystems möglichst genau zu erkennen.

4. Recherche nach Schwachstellen
   Aufgrund der gewonnenen Informationen aus der System- und Anwendungserkennung werden bekannte Schwachstellen für das Betriebssystem und die Dienste des Zielsystems recherchiert.

5. Ausnutzen der Schwachstellen
   Mit Hilfe gegebener Schwachstellen wird ein Angriffsversuch gestartet. Abschließend wird ein Sicherheitsreport über den Penetrations-Test erstellt.[14]

Die in dieser Diplomarbeit beschriebenen Fingerprinting-Ansätze werden zum Zweck der *System- und Anwendungserkennung* im Rahmen eines Penetrations-Tests angewandt. Durch verbesserte Fingerprinting-Methoden lassen sich schneller Sicherheitsprobleme in Netzwerken aufspüren. Dadurch steigt wiederum das allgemeine Sicherheitsniveau der IT-Infrastruktur von Unternehmen, die Penetrations-Tests durchführen lassen.

Bevor jedoch bestehende Fingerprinting-Ansätze besprochen werden können, sollen zunächst die benötigten netzwerktechnischen Grundlagen behandelt werden.

---

[13]  Vgl. BSI (2003)
[14]  Ebenda

## *2.2 Netzwerkgrundlagen, Schichtenmodelle und Protokolle*

### 2.2.1 Das TCP/IP-Referenzmodell

Für die erfolgreiche Kommunikation zwischen zwei Hosts müssen zunächst verschiedene Fragen beantwortet werden: Welche Art von Signal signalisiert eine 0 und welches Signal eine 1 auf dem Kommunikationsmedium? Welche Maßnahmen werden ergriffen, um Datenverluste durch Störeinflüsse (Dämpfung, Impendanz, Nebensprechen) zu verhindern? Wie findet eine Nachricht von Host A den Weg zu Host B, der sich möglicherweise auf der anderen Seite der Erdkugel befindet?

Diese Fragen werden durch Netzwerkprotokolle beantwortet. Ein Netzwerkprotokoll ist eine Vereinbarung, nach der Daten über ein Computernetzwerk ausgetauscht werden können. Protokolle beinhalten eine Syntax, also eine Spezifikation von Regeln und Formaten sowie deren zugeordnete Semantik (Bedeutung).

Die verschiedenen Aspekte einer Kommunikation werden in mehrere Schichten zerlegt und in einem Schichtenmodell dargestellt. Protokolle innerhalb einer Schicht erfüllen ähnliche Aufgaben und sind gegen Protokolle der gleichen Schicht austauschbar. Dieses Prinzip ermöglichte die Entwicklung des Internets, eines weltweiten Netzwerks verschiedenster Computersysteme mit verschiedensten Übertragungsmedien. Als Schichtenmodell kommt dabei das TCP/IP-Referenzmodell zum Einsatz, welches aus den folgenden Schichten besteht:

Schicht 4 – Anwendungsschicht
Die Anwendungsschicht enthält jene Netzwerkprotokolle, die von Anwendungen benutzt werden, mit denen der Endnutzer direkt in Berührung kommt, ob Browser, E-Mail-Kommunikation oder Dateitransfer.

Schicht 3 – Transportschicht
Die Transportschicht erstellt eine Ende-zu-Ende-Verbindung zwischen beiden Hosts und abstrahiert auf diese Weise von den darunter liegenden Schichten. Weiterhin adressiert die Transportschicht den jeweiligen Prozess auf Quell- und Zielhost über die Portnummer. Je nach Implementierung enthält die Transportschicht auch Flusskontroll-Mechanismen und Verfahren zur Segmentierung von Datenpaketen sowie Fehlerbehebungsverfahren, die den Erhalt sowie die korrekte Reihenfolge empfangener Daten sicherstellen.

Schicht 2 – Internetschicht
Die Vermittlungsschicht dient der Navigation durch den Netzgraphen, für die Kommunikation zwischen Computern, die sich in verschiedenen Netzen befinden. Sind die zu versendenden Pakete zu groß für die Sicherungsschicht, müssen sie von der Internetschicht fragmentiert werden.

Schicht 1 – Netzzugangsschicht
Die Netzzugangsschicht dient der Adressierung im lokalen Netzwerk, der Sicherstellung der Integrität übermittelter Daten sowie der Flusskontrolle, um Überlastungen des Netzmediums zu vermeiden. Außerdem spezifiziert die Netzzugangsschicht die physikalische Übertragung von Signalen zwischen zwei direkt verbundenen Computern über das verwendete Netzmedium.[15]

Soll Host A nun eine Nachricht an Host B versenden, so verpackt die Anwendungsschicht die Nutzlast in ein entsprechendes Format und reicht sie den Stapel hinunter. Die darunter liegenden Protokolle betrachten die erhaltene Informationseinheit nun als Nutzlast ihres eigenen Formates, fügen ihre eigenen Header (evtl. auch Trailer hinzu) und reichen die so erstellte Nachricht (Protocol Data Unit – PDU genannt) den Stapel weiter hinunter. Ist die Nachricht auf der Netzzugangsschicht angekommen, wird sie nach den Regeln des jeweiligen Netzmediums an Host B übertragen.

Ist die Nachricht bei Host B angekommen, so wird sie von diesem wieder „ausgepackt". Die Bitübertragungsschicht reicht die Nachricht an die Sicherungsschicht weiter, wonach die Nachricht stets an die nächsthöhere Schicht weitergereicht wird, bis sie auf der Anwendungsschicht angekommen ist. Damit ist der Übertragungsprozess abgeschlossen und die entsprechende Anwendung auf B kann die Nachricht darstellen.

Diese Beschreibung ist allerdings noch unscharf, denn, wenn sich A und B nicht im gleichen Netzwerk befinden, trifft die Nachricht auf ihrem Weg zu B auf eine gewisse

---

[15]   Vgl. Tanenbaum, Andrew S. (1992), S. 17ff

Zahl fremder Hosts. Diese fungieren als Router, d. h. sie leiten die Nachricht bis zur Vermittlungsschicht hoch, wo die Auswahl der besten Route erfolgt und schicken sie dann sogleich wieder den Protokollstapel bis zur Netzzugangsschicht herunter, um sie an den nächsten Host auf der Strecke zu senden.[16]

Nachdem der Aufbau des TCP/IP-Stacks deutlich geworden ist, müssen einige für das Fingerprinting relevante Netzwerkprotokolle genauer analysiert werden.

### 2.2.2 Das Internet-Protokoll in der Version 4 (IPv4)

Das Internet-Protokoll befindet sich im TCP/IP-Modell auf der Internetschicht. Es dient primär der Adressierung von Zielsystemen in lokalen TCP/IP-Netzen und im Internet. Für das Fingerprinting von Betriebssystemen spielt der Aufbau des IP-Headers eine besondere Rolle. Anhand verschiedener Implementierungen des TCP/IP-Stacks können Betriebssysteme voneinander unterschieden werden. Daher erfolgt ein Überblick über die Felder des IP-Headers:
Zunächst erfolgt eine Angabe der Versionsnummer (4 Bit). Im Folgenden soll nur die Version 4 des Internet-Protokolls betrachtet werden. Das nächste Feld speichert die Länge des IP-Headers (ebenfalls 4 Bit).

Es folgt das TOS-Feld (Type of Service, 8 Bit). Dieses Feld gibt an, welche Behandlung von dem jeweiligen Dienst bevorzugt wird: minimale Verzögerung, maximaler Durchsatz, maximale Zuverlässigkeit oder minimale Kosten. Auch wenn und gerade weil das TOS-Feld von den meisten TCP/IP-Implementierungen ignoriert wird, dient es der Unterscheidung verschiedener TCP/IP-Stacks.

Die ID eines Paketes (16 Bit) dient der Zusammensetzung fragmentierter IP-Pakete, wenn ein Paket zu groß für das verwendete Netzmedium ist. Die folgenden Flags (3 Bit) dienen der Konfiguration des Fragmentierungs-Verhaltens. Ist das More-Fragments-Bit gesetzt, so weiß das Zielsystem, dass weitere Fragmente des ursprünglichen Paketes folgen werden. Ist dagegen das Don't-Fragment-Bit gesetzt, so wird das Paket auf keinen Fall fragmentiert, mit der Konsequenz, dass es dann evtl. nicht zugestellt werden kann. Der Fragment-Offset (13 Bit) gibt schließlich an, welchen Bestandteil des ursprüngli-chen Paketes das gelieferte Fragment enthält, um das Paket in richtiger Reihenfolge wieder zusammensetzen zu können.

---

[16]  Vgl. Tanenbaum und Steen (2007), S. 117ff

Die Time To Live (TTL) gibt an, wie viele Stationen das IP-Paket auf seinem Weg zum Zielsystem maximal durchlaufen darf. Dadurch wird verhindert, dass Pakete auf ewig im Internet herumirren.

Ein weiteres Feld (8 Bit) gibt Auskunft über das verwendete Protokoll auf der Transportschicht. Darauf folgt eine Prüfsumme, um die Integrität des IP-Headers sicher zu stellen. Nun folgen IP-Adresse des Quell- und Zielsystems (jeweils 32 Bit) und mögliche Optionen, deren Implementierung jedoch freiwillig ist.[17]

### 2.2.3 Das Transmission Control Protocol

Das Transmission Control Protocol (TCP) ist unzweifelhaft einer der bedeutsamsten Erfindungen der Informationstechnik des 20. Jahrhunderts. TCP schafft es, auf einem fehlerbehafteten Medium auf Basis von unzuverlässigen Netzwerk-Protokollen eine zuverlässige Kommunikations-Verbindung zu etablieren, in der Nachrichten stets in der richtigen Reihenfolge ankommen und bei Störungen in den unteren Netzwerk-Schichten neu übertragen werden. Da die Felder des TCP-Headers, ebenso wie die Felder des IP-Headers, für das OS-Fingerprinting entscheidend sind, werden die wichtigsten Felder im Folgenden erläutert werden.

Die ersten 16 Bit werden für die Quell-Portnummer und die darauf folgenden 16 Bit für den Ziel-Portnummer einer Nachricht verwendet. Während die IP-Adresse das Zielsystem adressiert, identifiziert der Port den gewünschten Prozess auf dem Quell- und Zielsystem. Dabei werden von der Internet Assigned Numbers Authority (IANA) bestimmte Ports bestimmten Protokollen der Anwendungsschicht im TCP/IP-Modell zugeordnet. Horcht ein Server auf einer gegebenen Portnummer auf Anfragen und wird dieser Zugriff nicht durch etwaige Maßnahmen (Firewalls etc.) unterbunden, spricht man von einem offenen Port.[18]

---

[17] Vgl. Stevens, Richard W. (1994), S. 34
[18] Vgl. IANA (2008)

Die nächsten Felder im TCP-Header enthalten die Sequenznummer und die Bestätigungsnummer, jeweils 32 Bit. Den zu übertragenden Bytes wird eine fortlaufende Sequenznummer zugeordnet. Durch die entsprechende Bestätigungsnummer wird von der Gegenseite der Erhalt der Bytes bis zu genau jener Sequenznummer + 1 bestätigt. Dabei werden die Sequenznummern für jede Übertragungsrichtung getrennt verwaltet. Die initialen Sequenznummern müssen beim Verbindungsaufbau von beiden Seiten spezifiziert werden.[19]

Die Flags (6 Bit) spiegeln gewisse Protokoll-Zustände im Verlauf der Datenübertragung wieder. Sie werden z. B. zum Aufbau und Abbau von Verbindungen verwendet. Der Aufbau einer Kommunikationsverbindung erfolgt über ein dreistufiges Verfahren, dem sogenannten Three-Way-Handshake. Auslöser des Verbindungsaufbaus ist das gesetzte SYN-Flag im TCP-Header. Der Three-Way-Handshake[20] sei im Folgenden dargestellt:

1. Phase (Client): SYN, SEQn = X

2. Phase (Server): SYN, ACK, SEQn = Y, ACKn = X + 1

3. Phase (Client): ACK,  SEQn = X+1, ACKn = Y + 1

Legende:

SEQ, ACK = Gesetzte Flags im TCP-Header

SEQn        = Sequenznummer im TCP-Header

ACKn        =  Bestätigungsnummer im TCP-Header

Der Verbindungsabbau erfolgt auf ähnliche Weise über ein vierstufiges Verfahren und wird über das FIN-Flag eingeleitet.

Die Window Size (16 Bit) dient der Flusskontrolle. Sie legt fest, wie viele Bytes der Empfänger maximal in einem Puffer (Fenster) zwischenspeichern möchte. Um Überlastungen der Rechnerkapazitäten zu verhindern, gibt jede Seite der Gegenseite ihre bevorzugte Fenstergröße bekannt.[21]

---

[19]  Vgl. Stevens, Richard W. (1994), S. 223ff

[20]  Ebenda

[21]  Ebenda

Zuletzt folgen Optionen. Die Implementierung von Optionen in TCP/IP-Stacks ist unterschiedlich und zudem freiwillig. Auch die Implementierung von TCP-Optionen kann als Unterscheidungsmerkmal zum Zweck des OS-Fingerprintings dienen. Zu beachten ist, dass die Länge des Optionen-Feldes stets das Vielfache eines 32-Bit-Wortes sein muss.[22]

Mögliche Werte für das Feld Optionen sind:

- Window Scale (W)
  Ermöglicht deutlich größere Fenstergrößen als durch das Window-Size-Feld im TCP-Header spezifiziert werden können.

- No Operation (N)
  Wird verwendet, um die Länge des Optionen-Feldes bis zur Grenze eines 32-Bit-Wortes aufzufüllen.

- Maximum Segment Size (M)
  Bezeichnet die maximale Länge eines TCP-Segments.

- Timestamp (T)
  Die Timestamp-Option enthält einen 32-Bit-Timestamp. Unterscheiden sich Timestamps einer IP-Adresse, erlaubt das Rückschlüsse auf die Anzahl der Hosts hinter einem Masqueraded-NAT-Firewall. Ist das Betriebssystem bekannt, ermöglicht die Timestamp-Option außerdem, die Uptime eines Hosts zu schätzen.[23]

Nachdem die zentralen Eigenschaften von TCP/IP-Netzwerken erläutert wurden, ist der nächste Schritt die Anwendung dieses Wissens zum Zweck des Fingerprintings.

---

[22] Vgl. Firewall.cx (2008)
[23] Vgl. Arzur, Erwan (2005)

## *2.3 Analyse von Methoden des Fingerprintings*

### 2.3.1 Anwendung von Fingerprinting in der Informationstechnik

Der Fingerabdruck eines Menschen ist ein physisches Merkmal, das ihn von anderen Menschen unterscheiden lässt. In der Kriminalistik werden Fingerabdrücke vom Tatort gesammelt und mit bekannten Fingerabdrücken verglichen, um mögliche Täter ausfindig zu machen. Im übertragenen Sinne steht der Begriff Fingerabdruck auch für ein Identifizierungsmerkmal im Sinne einer einzigartigen Eigenschaft, die Personen oder Dingen zugeordnet wird. Dies findet seinen Ausdruck in den Begriffen „genetischer Fingerabdruck" oder „digitaler Fingerabdruck"[24].

In der Informationstechnik finden sich die Begriffe des OS- und Service-Fingerprintings. Gemeint ist eine Zusammenfassung gewisser Verhaltensweisen, durch die verschiedene Betriebssysteme oder Dienste zuverlässig voneinander unterschieden werden können. Solch ein Fingerprint eines Betriebssystems könnte folgendermaßen aussehen:

```
FingerPrint  IRIX 6.2 - 6.4
TSeq(Class=i800)
T1(DF=N%W=C000|EF2A%ACK=S++%Flags=AS%Ops=MNWNNT)
T2(Resp=Y%DF=N%W=0%ACK=S%Flags=AR%Ops=)
T3(Resp=Y%DF=N%W=C000|EF2A%ACK=O%Flags=A%Ops=NNT)
T4(DF=N%W=0%ACK=O%Flags=R%Ops=)
T5(DF=N%W=0%ACK=S++%Flags=AR%Ops=)
T6(DF=N%W=0%ACK=O%Flags=R%Ops=)
T7(DF=N%W=0%ACK=S%Flags=AR%Ops=)
PU(DF=N%TOS=0%IPLEN=38%RIPTL=148%RID=E%RIPCK=E%UCK=E
%ULEN=134%DAT=E)
```

Quelle: Vaskovich, Fyodor (1999)

Die Semantik dieser Darstellung eines Fingerprints soll im folgenden Kapitel entschlüsselt werden.

---

[24] Gemeint sind hier elektronische Signaturen.

### 2.3.2 OS-Fingerprinting durch TCP/IP-Stack-Analyse

Die Erkennung entfernter Betriebssysteme durch OS-Fingerprinting geschieht für gewöhnlich durch eine Analyse des TCP/IP-Stacks. Dabei werden zunächst verschiedene, speziell vorbereitete Nachrichten an das Zielsystem gesendet. Anhand des unterschiedlichen Antwort-Verhaltens verschiedener TCP/IP-Implementierungen in verschiedenen Betriebssystemen können Rückschlüsse auf das verwendete Betriebssystem gezogen werden. Durch Kombination verschiedener Tests lässt sich die Treffer-Wahrscheinlichkeit verbessern.[25]

Oben gezeigter Fingerprint gehört zu einem Betriebssystem, das einer TCP/IP-Stack-Analyse unterzogen wurde. Er wurde durch das Netzwerk-Analyse-Tool Nmap (Network Mapper) erzeugt. Im Folgenden soll die Bedeutung jenes Fingerprint-Formates geklärt werden.

Die 1. Zeile des Fingerprints gibt den Gültigkeitsbereich des Fingerprints an. Oben gezeigter Fingerprint deckt die Versionen 6.2 - 6.4 des Betriebssystems IRIX ab.

Die folgenden Zeilen stellen verschiedene Tests auf den TCP/IP-Stack und deren spezifischen Ergebnisse dar. Zunächst erfolgt ein Test auf die Erzeugung der TCP-Sequenznummern, der TSeq. Die folgenden Tests beinhalten verschiedene Szenarien des TCP-Einsatzes und werden von T1 – T7 durchnummeriert. Der letzte Test, PU, stellt den sogenannten Port Unreachable Test dar.[26]

Die Bedeutung dieser Tests sollen in Tabelle 1 dargelegt werden.

---

[25] Ein Administrator kann sich teilweise vor OS-Fingerprinting schützen. Smart et al. (2000) beschreiben eine Methode, eine TCP/IP-Stack-Analyse durch Umkonfiguration der TCP/IP-Parameter im Betriebssystem zu erschweren.

[26] Vgl. Vaskovich, Fyodor (1999)

Tabelle 1: Testverfahren im Rahmen einer TCP/IP-Stack-Analyse

| Test | Bedeutung |
|------|-----------|
| TSeq | Eine Serie von Paketen mit gesetztem SYN-Flag wird an das Zielsystem versandt, um zu testen, wie TCP-Sequenz-Nummern vom TCP/IP-Stack des Zielsystems berechnet werden. |
| T1 | Ein Paket mit gesetztem SYN-Flag und gesetzten Optionen (W, N, M, T) wird an einen offenen TCP-Port versandt. |
| T2 | Ein Paket ohne gesetzte TCP-Flags und mit gesetzten Optionen (W, N, M, T) wird an einen offenen TCP-Port versandt. |
| T3 | Ein Paket mit gesetzten TCP-Flags (SYN, FIN, PSH, URG) und gesetzten Optionen (W, N, M, T) wird an einen offenen TCP-Port versandt. |
| T4 | Ein Paket mit gesetztem ACK-Flag und gesetzten Optionen (W, N, M, T) wird an einen offenen TCP-Port versandt. |
| T5 | Ein Paket mit gesetztem SYN-Flag und gesetzten Optionen (W, N, M, T) wird an einen geschlossenen TCP-Port versandt. |
| T6 | Ein Paket mit gesetztem ACK-Flag und gesetzten Optionen (W, N, M, T) wird an einen geschlossenen TCP-Port versandt. |
| T7 | Ein Paket mit diversen gesetzten Flags (FIN, PSH, URG) und gesetzten Optionen (W, N, M, T) wird an einen geschlossenen TCP-Port versandt. |
| PU | Ein Paket wird an einen geschlossenen UDP-Port versandt. |

Quelle: Vgl. Glaser, Thomas (2000)

Für die Auswertung dieser Tests werden verschiedene Metriken definiert, die das unterschiedliche Antwort-Verhalten dokumentieren sollen. Im Fingerprint werden diese Metriken mit ihrem entsprechenden Antwort-Wert festgehalten. Metriken werden untereinander durch das Prozent-Zeichen (%) getrennt. Die Pipe (|) dient dabei als logisches Oder. Fehlt eine Metrik im Fingerprint, so kam sie in der Antwort-Nachricht auf diesen Test nicht vor.

Diese Metriken werden in Tabelle 2 beschrieben.

Tabelle 2: Metriken für die Auswertung einer TCP/IP-Stack-Analyse

| Metrik | Wertebereich | Beschreibung |
|--------|-------------|--------------|
| Resp | Y = Es gab eine Antwort<br>N = Es gab keine Antwort | Reaktion des Zielsystems auf eine Anfrage-Nachricht |
| DF | Y = DF-Bit war gesetzt<br>N = DF-Bit war gesetzt | Wert des „Don't Fragment"-Bits |
| W | 16-Bit-Integer | Gewünschte Fenstergröße in der Antwort-Nachricht |
| ACK | 0 = ACK mit Sequenznummer 0<br>S = ACK-Sequenznummer<br>S++ = ACK-Sequenznummer + 1 | Bestätigungsnummer in der Nachricht |
| Flags | S = SYN<br>A = ACK<br>R = RST<br>F = FIN<br>U = URG<br>P = PSH | Gesetzte TCP-Flags waren in der Antwort-Nachricht |
| Ops | W = Window Scale<br>N = No Operation<br>M = Max. Segmentation Size<br>E = Echoed MSS<br>T = Timestamp | TCP-Optionen in der Antwort-Nachricht. Auch die Reihenfolge der gesetzten Optionen wird als Unterscheidungsmerkmal herangezogen. |
| TOS | 8-Bit-Integer | Type-Of-Service-Feld des IP-Headers |

Quelle: Vgl. Glaser, Thomas (2000)

Diese Tests und Metriken bieten eine Vielfalt an Unterscheidungsmöglichkeiten für TCP/IP-Implementierungen und ermöglichen eine grundlegende Differenzierung verschiedener Betriebssysteme. Verbesserungspotentiale bedarf jedoch noch die Feinerkennung unterschiedlicher Releases des gleichen Betriebssystems.[27]

---

[27]  Vgl. Poppa, Ryan (2007)

Wie jedoch könnte die Zuverlässigkeit des OS-Fingerprintings verbessert werden? Zu diesem Zweck gilt es, neue Tests und Metriken zu finden, die weitere Unterschiede in der Implementierung eines TCP/IP-Stacks erfassen. Neue Tests und Metriken könnten durch die Untersuchung von Changelogs oder durch die Übernahme aus bestehender Fingerprinting-Software gewonnen werden.

Durch die Anwendung von *Fuzzing* könnte die Erzeugung von neuen Testfällen automatisiert werden. Zu diesem Zweck entwickelt der Verfasser ein eigenes Konzept für OS-Fingerprinting, dass in Kapitel 3 beschrieben ist.

Problematisch ist auch, dass oben beschriebene Tests und Metriken im Quellcode von Nmap hart kodiert und somit eher schwer zu verändern sind. Deshalb wird der Verfasser für die Implementierung seines Konzeptes eine eigene OS-Fingerprinting-Software entwickeln, die in Kapitel 4 beschrieben wird.

Während sich das OS-Fingerprinting in erster Linie auf der Transportschicht und der Internetschicht abspielt, nutzt das Service-Fingerprinting auch Unterschiede zwischen Implementierungen von Protokollen der Anwendungsschicht aus. Bekannte Ansätze des Service-Fingerprintings sollen im Folgenden dargestellt werden.

### 2.3.3 Port-Scanning zur Vorbereitung von Service-Fingerprinting

Die zentrale Frage, die ein Service-Fingerprinting beantworten soll, ist: Welche Dienste laufen auf dem Zielsystem?

Vor dem eigentlichen Fingerprinting folgt üblicherweise ein Port Scan, um offene Ports, hinter denen sich die zu analysierenden Dienste verbergen, aufzuspüren. Der Port Scan geschieht im einfachsten Fall durch einen TCP-Verbindungsaufbau und den anschließenden Abbau (Full-Connect-Scan). War der Verbindungsaufbau erfolgreich, so ist der getestete Port offen und die Verbindung wird wieder abgebaut. Eine effizientere Variante, vor allem wenn viele Ports in einem Durchlauf untersucht werden sollen, ist der Half-Open-Scan. Dabei erfolgen die ersten zwei Schritte des Three-Way-Handshakes in üblicher Weise, doch im dritten Schritt wird statt der Bestätigung der Kommunikation durch den Client der Verbindungsaufbau durch eine Nachricht mit gesetztem RST-Flag abrupt beendet. Auf diese Weise benötigt der Scan eines Ports lediglich 3 Nachrichten (Half-Open-Scan) statt 7 Nachrichten (Full-Connect-Scan).[28]

Der oben beschriebene Algorithmus für einen Half-Open-Scan würde in Pseudocode[29] formuliert folgendermaßen aussehen:

```
find-open-tcp-port (n-ports):

for port-no := 0 to n-ports

    socket := create-socket ()
    synchronize-connection (socket)

    if ( connection-acknowledged (socket) )
        reset-connection (socket)
        return port-no
    end-if

end-for

return NO_PORT_FOUND
```

Der Aufruf eines Half-Open-Scans mit dem Port-Scanner Nmap (Network Mapper) sieht folgendermaßen aus:

```
~ # nmap -sS scanme.nmap.org

Starting nmap 3.83.DC13 ( http://www.insecure.org/nmap/ ) at 2008-07-06 04:02 CEST
Interesting ports on scanme.nmap.org (64.13.134.52):
(The 1661 ports scanned but not shown below are in state: filtered)
PORT   STATE  SERVICE
22/tcp  open   ssh
25/tcp  closed smtp
53/tcp  open   domain
70/tcp  closed gopher
```

---

[28]  Vgl. Ruef, Marc (2007), S. 327ff
[29]  Vgl. Anhang 1

80/tcp  open   http

113/tcp closed auth

Nmap finished: 1 IP address (1 host up) scanned in 37.019 seconds

Mit Hilfe dieser Information kann nun begonnen werden, die Prozesse hinter diesen Ports zu analysieren.

### 2.3.4 Application-Mapping und Banner-Grabbing

Der Ansatz des Application-Mappings, Spekulationen über das Protokoll auf TCP/IP-Anwendungsschicht hinter einem offenen Port zu betreiben, ist der Vergleich mit offiziellen Port-Mappings.[30] Im Bereich von 0 – 49151 registriert die IANA Portnummern für bestimmte Anwendungen.[31] So ist es sehr wahrscheinlich, dass sich hinter Port 80 ein Webserver befindet, denn gängige Web-Browser verwenden standardmäßig diesen Port, um sich zu einem Webserver zu verbinden. Die Anwendung von Application Mapping hat vermutlich hohe Erfolgsquoten, kann aber leicht ausgehebelt werden. Ein Netzwerk-Administrator könnte etwa bestimmte Dienste absichtlich auf unerwarteten Portnummern horchen lassen, um mögliche Angreifer zu verwirren.

Vor allem aber lässt sich auf diese Weise nur das zugrunde liegende Anwendungs-Protokoll erkennen, nicht jedoch ein konkreter Dienst oder gar ein bestimmtes Release dieses Dienstes. Dies sollte jedoch das Ziel eines erfolgreichen Fingerprintings sein, um den passenden Exploit für den jeweiligen Dienst auswählen zu können. Schließlich ist dieser Ansatz untauglich, wenn sich hinter den geöffneten Ports nicht bei der IANA registrierte Dienste und insbesondere Individualsoftware-Produkte verbergen.[32]

Ein anderer Ansatz für die Analyse von Diensten ist das sogenannte Banner-Grabbing. Dabei wird gezielt eine TCP-Verbindung zu einem offenen Port des Zielsystems aufgebaut, um den Dienst dazu zu bewegen, eine Willkommensmeldung (Banner) auszuge-ben, in der er sich selbst zu erkennen gibt. Folgendes Beispiel soll die Wirksamkeit eines gezielten Banner-Grabbings verdeutlichen:

```
~ $ telnet hackthissite.org 80
Trying 207.210.114.39...
Connected to hackthissite.org.
Escape character is '^]'.
GET /index.html HTTP/1.1
Host: www.example.net

HTTP/1.1 301 Moved Permanently
Date: Sun, 06 Jul 2008 03:05:02 GMT
Server: Apache/2.2.4 (FreeBSD) mod_ssl/2.2.4 OpenSSL/0.9.8e DAV/2 PHP/5.2.3
with Suhosin-Patch mod_perl/2.0.3 Perl/v5.8.8 [...]
```

---

[30]  Auch Nmap benutzt diese Methode, um die Protokoll-Zuordnungen in obiger Ausgabe zu erstellen.
[31]  Vgl. IANA (2008), Dieses Port-Mapping wird meist zum Betriebssystem mitgeliefert. In gängigen UNIX-Derivaten ist es etwa in der Datei /etc/services abrufbar.
[32]  Vgl. Ruef, Marc (2007), S. 395ff

Dieser Webserver verrät nicht nur die eingesetzte Server-Software, sondern auch Details über die spezifische Version, verwendete Module, Patches und das zugrunde liegende Betriebssystem. Dieses Beispiel macht auch deutlich, dass Banner-Grabbing ebenfalls zur Erkennung des Betriebssystems eingesetzt werden kann.

Inzwischen sind viele Server-Betreiber dazu übergegangen, die Ausgabe von Bannern soweit wie möglich zu beschränken. Damit sinkt die Erfolgsquote von Banner-Grabbing beträchtlich und macht den Einsatz fortgeschrittener Methoden erforderlich.[33]

### 2.3.5 Methoden des Service-Fingerprintings

Ziele des Service-Fingerprintings sind[34]:

1. Das zugrunde liegende Anwendungsprotokoll zu ermitteln

2. Die verwendete Server-Software herauszufinden

3. Das genaue Release zu identifizieren

---

[33] Vgl. Arkin, Ofir (2003)
[34] Vgl. Ruef, Marc (2007), S. 492

Dafür müssen wie bei der TCP/IP-Stack-Analyse entsprechende Testfälle erstellt, über das Netzwerk versandt und die Antworten-Nachricht gesammelt und analysiert werden.

Service-Fingerprinting-Software wird meist unter der Annahme entwickelt, dass das Anwendungsprotokoll bereits bekannt ist. Da jedes Protokoll der Anwendungsschicht ein anderes Vorgehen erfordert, kann es keine universelle Service-Fingerprinting-Software geben, es sei denn, als Zusammenstellung verschiedener Fingerprinting-Ansätze.

Mit der Kenntnis des Anwendungsprotokolls können speziell präparierte Nachrichten an das Zielsystem und den Zielport gesendet werden. Diese Nachrichten zeichnen sich dadurch aus, dass sie das jeweilige Anwendungsprotokoll grundsätzlich einhalten, jedoch gewisse (dem Protokoll konforme oder vom Protokoll abweichende) Änderungen vornehmen, die zu einem unterschiedlichen Antwort-Verhalten der Zielanwendungen führen. Aufgrund dieser Unterschiede im Antwort-Verhalten können daraufhin Fingerprints der jeweiligen Dienste erstellt werden, die zur späteren Identifikation des gleichen oder eines ähnlichen Dienstes dienen.[35]

Im Folgenden wird ein Beispiel für das Fingerprinting von Webservern erklärt. Als verwendete HTTP-Methode kommt dabei HEAD zum Tragen, die lediglich den Header der HTTP-Antwort zurück liefert. Das hat den Grund, dass der Content der angeforderten URL nicht zur Unterscheidung dient. Schließlich können mit dem gleichen Webserver-Release verschiedene Inhalte bereitgestellt werden. Ferner wird eine HTTP-Anfrage mit überlangem URL erzeugt und über das Netzwerk versendet. Diese Anfrage könnte z. B. folgendermaßen aussehen:

> HEAD /aMp7xMqT[...] HTTP/1.1
> Host: example.org

Die Antwort des Webservers enthält u. a. den HTTP-Statuscode, der verschiedene Situationen, die in HTTP auftreten können, charakterisiert. Für den Fall einer überlangen Antwort wären verschiedene Statuscodes denkbar: 404 Not Found, 403 Forbidden, und 414 Request-URI Too Large. Dabei werden z. B. vom Webserver Apache, je nach Re-lease und Länge des übergebenen URLs, verschiedene Statuscodes zurückgeliefert, die zur Identifizierung des Apache-Releases dienen können.

---

[35]  Vgl. Ruef, Marc (2007), S. 495ff

Wie von Ruef gezeigt wird, bewirken nicht nur unterschiedliche Releases von Apache, sondern auch das verwendete Betriebssystem einen Unterschied. Dieser Test ist also gleichermaßen zur Erkennung des Apache-Releases wie auch zur Erkennung des Betriebssystems nützlich. Aktuelle Releases des Apache scheinen jedoch stets die gleichen Statuscodes für eine gegebene URL-Länge zu verwenden.[36]

Da sich das Vorgehen des Service-Fingerprintings je nach Art des Dienstes unterscheidet, scheint die systematische Erzeugung von Testfällen mittels *Fuzzing* für die Anwendung im Service-Fingerprinting besonders fruchtbar. So kann nicht nur die Genauigkeit bestehender Service-Fingerprinting-Software verbessert werden sondern auch die Entwicklung von Fingerprinting-Software für neue Anwendungsprotokolle erleichtert werden. Tatsächlich erlaubt die Genauigkeit bestehender Service-Fingerprinting-Software noch Spielraum für Verbesserungen. So gelingt es dem Verfasser wie in Kapitel 5 beschrieben *mittels Anwendung von Fuzzing* zwei Dienste voneinander zu unterscheiden, die von bestehender Service-Fingerprinting-Software noch nicht unterschieden werden können. Bekannte Methoden des Fuzzings sollen in Kapitel 2.5 erläutert werden.

## 2.4 Verwandte Arbeiten

Verwandte Arbeiten im Bereich OS- und Service-Fingerprinting sollen im Folgenden kurz dargestellt und vom Thema dieser Arbeit abgegrenzt werden.

Lippmann et al. präsentieren eine Studie über passives OS-Fingerprinting, bei der systematisch verschiedene Klassen von Betriebssystemen gebildet werden.[37] Im Gegensatz hierzu beschäftigt sich die vorliegende Arbeit mit aktivem Fingerprinting, d.h. es werden initiativ Anfrage-Nachrichten erzeugt und versendet, um die zugehörigen Antwort-Nachrichten systematisch zu analysieren.

---

[36] Vgl. Ruef, Marc (2007), S. 533
[37] Vgl. Lippmann et al. (2003)

Sarraute und Burroni berichten über ein fortgeschrittenes OS-Fingerprinting-Verfahren, dass auf neuronalen Netzwerken basiert[38]. Im Gegensatz zu der vorliegenden Arbeit verwenden Sarraute und Burroni jedoch keine systematisch erzeugte Anfragemenge. Durch die systematische Erzeugung von Anfrage-Nachrichten mittels Fuzzing können prinzipiell beliebige Unterschiede im Antwort-Verhalten von Betriebssystemen und Diensten erkannt werden.[39]

Gagnon et al. schlagen eine hybride Vorgehensweise bzgl. des OS-Fingerprintings vor, die passives und aktives Fingerprinting miteinander verbindet.[40] Dabei verwenden Gagnon et al. wissensbasierte Systeme, die nur dann aktive Fingerprinting-Verfahren einsetzen, wenn dieses notwendig ist, um auf der Basis bereits vorhandenen Informationen eine weitere Eingrenzung vorzunehmen. Im Unterschied dazu werden in der vorliegenden Arbeit nur aktive Fingerprinting-Mechanismen verwendet und es wird bewusst keine Einschränkung der Anfrage-Menge vorgenommen, um möglichst viele verschiedene Betriebssysteme und Dienste voneinander abgrenzen zu können.

Caballero et al. beschreiben eine erste Studie über Fingerprinting auf Basis automatisierter Anfrage-Nachrichten.[41] Anders als in der vorliegenden Arbeit werden dabei Methoden des Machine Learning eingesetzt. Der Einsatz von Fuzzing hat gegenüber Machine Learning den Vorteil, das Anfrage-Antwort-Verhalten verschiedener Systeme (auch zukünftiger) möglichst detailliert abzubilden. Der Nachteil ist jedoch, dass erheblich mehr Anfrage-Nachrichten versendet werden müssen.

Greenwald und Thomas beschreiben, wie man den Netzwerkverkehr eines aktiven Fingerprinting-Verfahren verbergen kann.[42] Im Unterschied zu der vorliegenden Arbeit werden nur 1 bis 3 Anfrage-Nachrichten zum Fingerprinting verwendet. Während diese Maßnahmen die Detektion des Fingerprintings erschweren können, sind diese jedoch nicht mit einer systematischen Analyse auf Basis von großen Mengen automatisiert erzeugter Anfrage-Nachrichten vereinbar. Allerdings wäre es durchaus denkbar, auf Basis der hier durchgeführten Analyse eine kleine Menge spezifischer Anfrage-Nachrichten auszuwählen, die für das OS- und Service-Fingerprinting von besonderer Relevanz ist.

---

[38] Vgl. Sarraute und Burroni (2005)
[39] Vgl. Kapitel 3
[40] Vgl. Gagnon et al. (2007)
[41] Vgl. Caballero et al. (2007)
[42] Vgl. Greenwald und Thomas (2007)

## *2.5 Analyse von Methoden des Fuzzings*

### 2.5.1 Anwendung von Fuzzing für systematische Schwachstellentests

Der Begriff *Fuzzing* bezeichnet eine Technik zur systematischen Fehler- und Schwachstellensuche von Formaten, Programmen und Protokollen. Die Idee stammt von Professor Barton Miller, der sie 1989 als Projektarbeit für seine Studenten an der University of Winsconsin-Madison herausgab. Ziel des Projekts war es, die Robustheit von UNIX-Programmen mit einem unvorhersehbaren Input-Strom zu testen.

Das Projekt gliederte sich in folgende Schritte:[43]

1. Zuerst sollte ein sogenannter Fuzz-Generator programmiert werden. Dies ist ein Programm, welches einen zufälligen Zeichenstrom generiert.
2. Im zweiten Schritt sollten möglichst viele UNIX-Utilities mit dem Output des Fuzz-Generators gefüttert werden, mit dem Ziel, diese zum Absturz zu bringen.
3. Im Falle eines Absturzes, sollten die spezifischen Zeichenketten, die das Programm zum Absturz bringen, ermittelt werden.
4. Zuletzt sollte der genaue Grund für den jeweiligen Programmabsturz ermittelt werden, um den Fehler beheben zu können.

Im Verlauf des Projektes konnten mehr als 24% der getesteten Programme zum Absturz gebracht werden, bei einer Menge von 90 Programmen unter dem Einsatz von 7 verschiedenen UNIX-Versionen.[44]

Auch heute noch ist Fuzzing ein wichtiges Testverfahren für sicherheitsbewusste Entwickler. Im Juli 2006 verkündete H. D. Moore den „Month of Browser Bugs". Ziel war es, jeden Tag des Monats einen sicherheitsrelevanten Browser-Fehler zu entdecken, was ihm mittels Anwendung von Fuzzing auch tatsächlich gelang.[45] Vor diesem Hintergrund lohnt es sich, einen genaueren Blick auf den Fuzzing-Prozess[46] zu werfen.

---

[43]  Vgl. Miller, Barton P. (1989)
[44]  Ebenda
[45]  Vgl. Sutton et al. (2007), S. 268
[46]  a. a. O, S. 29

### 2.5.2 Der Fuzzing-Prozess

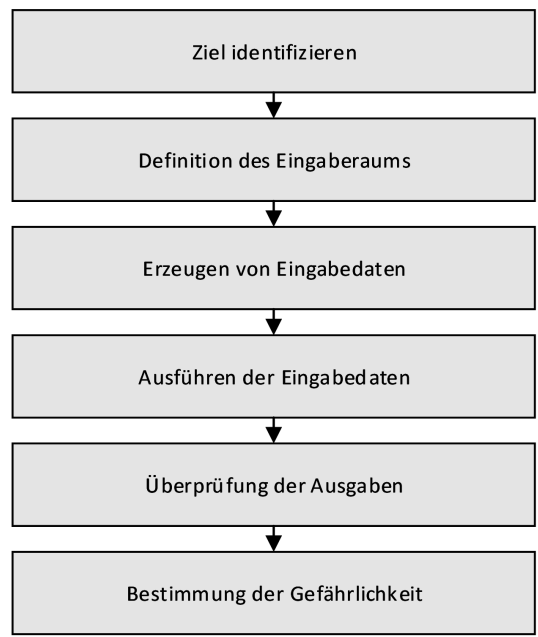

Abbildung 1: Der Fuzzing-Prozess

Der Fuzzing-Prozess (Abbildung 1) ist ein 6-stufiges Phasenmodell, das das Vorgehen beim Einsatz von Fuzzing zum Zweck des Software-Testing beschreibt. Die einzelnen Schritte[47] werden im Folgenden näher erläutert.

1. Ziel identifizieren

   Um eine qualifizierte Entscheidung über die Auswahl der Fuzzing-Methoden und Tools treffen zu können, müssen zunächst Informationen über das Ziel gesammelt werden. Handelt es sich um eine Datei, ein Programm oder um einen entferntes System? Welche Protokolle liegen dem Ziel zugrunde?

2. Definition des Eingaberaums

   Im nächsten Schritt werden die möglichen Eingabeformen des Ziels (Dateien, Argumente, Nachrichten aus dem Netzwerk) identifiziert, um den Ansatzpunkt für eine erfolgreiche Schwachstellensuche zu finden. Die Wahl der Angriffspunkte ist immer ein Kompromiss zwischen Laufzeit und Vollständigkeit. Es ist hilfreich, die Laufzeit im Vorhinein zu überschlagen, schließlich kann nicht bewiesen werden, dass das getestete Programm P mit den Eingabedaten D überhaupt terminiert. Diese Problematik ist auch als das Halteproblem bekannt.[48]

---

[47]  Vgl. Sutton et al. (2007), S. 27f

[48]  Vgl. Asteroth und Christel (2003), S. 99ff

3. Erzeugen von Eingabedaten

   Das Erzeugen der präparierten Eingabedaten (Fuzz) geschieht über unterschiedliche Fuzzing-Methoden (vgl. Kapitel 2.5.3 - 2.5.6). Es können völlig zufällige Daten generiert, aber auch gültige Eingaben verändert werden. Mit einer gegebenen Fuzzing-Methode kann die Erzeugung der Testfälle automatisiert werden.

4. Ausführen der Eingabedaten

   Nun kann das Programm, das Format oder das Protokoll mit den erzeugten Eingabedaten aufgerufen werden. Dieser Prozess kann ebenfalls automatisiert werden, was den Einsatz von Fuzzing gegenüber anderen Methoden des Software-Testing (etwa. statische Analyse oder formale Verifikation) besonders attraktiv macht.

5. Überprüfung der Ausgaben

   Nach Aufruf des Programms müssen die Ausgaben auf etwaige Hinweise auf Schwachstellen untersucht werden. Dieser Vorgang ist je nach eingesetzter Fuzzing-Software sehr unterschiedlich und enthält zugleich automatisierbare und manuelle Schritte.

6. Bestimmung der Gefährlichkeit

   Schließlich gilt es, die gefundenen Schwachstellen zu kategorisieren und Möglichkeiten, das Zielsystem zu kompromittieren zu erforschen. Die Entwicklung von Exploits kann durch Exploit-Frameworks wie Metasploit teilweise automatisiert werden.[49]

Für das Fingerprinting steht jedoch nicht die Suche nach Schwachstellen im Blickwinkel der Betrachtung. Vielmehr gilt es das *Erzeugen von Eingabedaten* für eine umfassendere Verhaltens-Analyse im Vergleich zu bestehenden Fingerprinting-Tools zu nutzen. Dabei ist die Wahl der richtigen Fuzzing-Methode von großer Bedeutung. Verschiedene Fuzzing-Methoden werden im Folgenden vorgestellt.

---

[49] Vgl. Singh und Mookhey (2004)

### 2.5.3 Wahlloses Fuzzing

Bei wahllosem Fuzzing wird die Ziel-Anwendung mit beliebigen zufälligen Daten gefüttert. Da der Eingaberaum nach oben unbeschränkt ist, ist die Laufzeit beim Einsatz von wahllosem Fuzzing unbestimmt. Eine sinnvolle Testlänge ist schwer zu bestimmen, denn möglicherweise treten Fehler erst bei „besonders langen" Eingabedaten auf. Außerdem enthalten die meisten Programme diverse Plausibilitätschecks, die eine völlig wahllose Eingabemenge im Regelfall verwerfen. Im Allgemeinen ist wahlloses Fuzzing also eine eher ineffiziente Methode zur Schwachstellensuche und zeigt die Notwendigkeit effizienterer Methoden auf.[50]

### 2.5.4 Mutation-Based-Fuzzing

Der Ansatz des Mutation-Based-Fuzzing setzt zunächst gültige Eingabemengen voraus. Diese Eingabemengen „mutieren" anschließend wiederholt, um ein unerwartetes Verhalten zu produzieren. Somit kommen Plausibilitätschecks erst bei stärker mutierten Daten zum Tragen und die Wahrscheinlichkeit steigt, schneller auf Schwachstellen zu stoßen. Problematisch bei diesem Ansatz ist, dass der Eingaberaum nach wie vor nach oben unbeschränkt und somit die Laufzeit unbestimmt ist.[51]

Die Anwendung von Mutation-Based-Fuzzing eignet sich aus Sicht des Verfassers besonders für Protokolle und Formate mit variablen Feldgrößen, deren Wertebereich ohnehin nach oben unbeschränkt ist. Beispiele hierfür sind Protokolle der Anwendungsschicht wie FTP und HTTP.

### 2.5.5 Generation-Based-Fuzzing

Der Ansatz des Generation-Based-Fuzzing setzt auf bekannten Protokoll-Spezifikationen auf. Dabei versucht der Entwickler des Fuzzers, ein Template zu erzeugen, dass die Struktur eines Protokolls beschreibt.[52]

Die Anwendung von Generation-Based-Fuzzing ist aus Sicht des Verfassers besonders für Formate und Protokolle mit festen Feldgrößen sinnvoll: Da mögliche Wertebereiche nach oben beschränkt sind, können in endlicher Zeit alle mögliche Wertekombinationen getestet werden. Erscheint die dafür benötigte Zeit dennoch zu hoch, so kann eine Auswahl des maximalen Wertebereichs erzeugt werden, indem für jedes Feld jeweils der n-te mögliche Wert getestet wird. Typische Anwendungsfälle sind Protokolle der unteren Schichten des TCP/IP-Stacks: Ethernet, IP, TCP.

---

[50]  Vgl. Rütten, Christiane (2007)
[51]  Ebenda
[52]  Vgl. Miller, Charlie (2007), S. 7 sowie Sutton et al. (2007), S. 36

### 2.5.6 Evolutionary Fuzzing

Der Ansatz des Evolutionary Fuzzings ist eine konsequente Weiterentwicklung des Mutation-Based-Fuzzings und bedient sich genetischer Algorithmen, also Strategien aus der Evolutionstheorie, um die Erfolgsquote von Testfällen zu verbessern. Neben der Mutation von Testfällen umfasst dies auch die Strategien:

- Selektion: Die *fitness* des Testfalls bestimmt, ob dieser in eine spätere Generation übernommen werden soll.
- Rekombination: Eigenheiten mehrerer Testfälle werden zu einem neuen Testfall verknüpft.[53]

Ein Problem genetischer Algorithmen ist die Definition einer passenden Metrik zur Bestimmung der *fitness* eines Testfall. Für den Einsatz von Evolutionary Fuzzing bekannte Metriken sind:

- Criticality: Die Anzahl mit einem Testfall abgedeckter potentiell gefährlicher Bibliotheksfunktionen im Quellcode.
- Code Coverage: Das Ausmaß der Abdeckung des Quellcodes durch einen Testfall.[54]

Für ein Fingerprinting auf Basis von Evolutionary Fuzzing müsste zunächst eine geeignete Metrik für die *fitness* von Testfällen definiert werden und auf Basis dieser Metrik verschiedene Generationen von Testfällen erzeugt werden. Auf diese Weise kann die Zeit zur Unterscheidung von Zielsystemen gegenüber dem Ansatz des Mutation-Based-Fuzzing möglicherweise verringert werden. Die Anwendung von Evolutionary Fuzzing würde jedoch den zeitlichen Rahmen dieser Diplomarbeit sprengen und wird deshalb nicht weiter betrachtet.

---

[53]  Vgl. Engesser et al. (1993), S. 269
[54]  Vgl. Miller, Charlie (2007), S. 7

Auf Grundlage des Fuzzing-Prozesses und der bis hier erörterten Grundlagen soll im Folgenden ein eigenes Konzept für eine Fingerprinting-Software erstellt werden.

# 3 Konzept eines Fingerprinting mittels Anwendung von Fuzzing

## 3.1 Ansatz für ein Fingerprinting auf Basis von Fuzzing

Gegeben sei ein Computersystem (im Folgenden *Quellsystem* genannt), auf dem die hier konzipierte Fingerprinting-Software läuft. Zum Zweck des Fingerprintings werden Zielsysteme analysiert.

Mögliche *Zielsysteme* sind alle Computersysteme, die über TCP/IP mit dem Quellsystem verbunden sind und mindestens über ihre IP-Adresse bzw. ihren Hostnamen bekannt sind.

Eine *Anfrage-Nachricht* (auch: Testfall) ist eine PDU die vom Quellsystem an das Zielsystem versandt wird. Zweck ist es, verschiedene Antwort-Nachrichten auf verschiedenen Zielsystemen zu provozieren. Das für die Anfrage-Nachricht verwendete Protokoll ist abhängig davon, ob Betriebssysteme oder Dienste eines bestimmten Protokolls erkannt werden sollen.

Eine *Antwort-Nachricht* ist eine PDU, die von einem Zielsystem als Reaktion auf eine empfangene Anfrage-Nachricht an das Quellsystem versandt wird.

Diese Testumgebung sei in Abbildung 2 dargestellt.

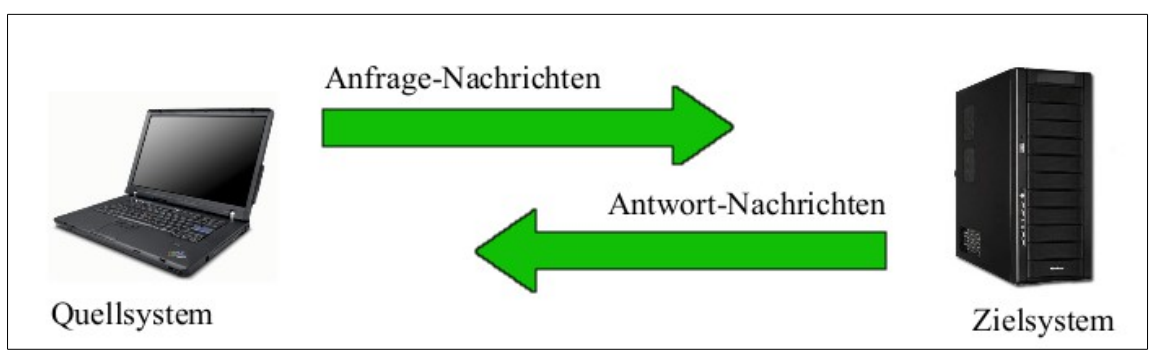

Abbildung 2: Versuchsaufbau

Die Analyse des Antwort-Verhaltens eines Zielsystems setzt eine *Verkettung* von An­frage-Nachrichten und Antwort-Nachrichten voraus. Das bedeutet, dass von einer emp­fangenen Antwort-Nachricht stets auf die versendete Anfrage-Nachricht geschlossen werden kann.

Das *Antwort-Verhalten* eines Zielsystems ist die Abbildung der Menge möglicher An­frage-Nachrichten (*Eingaberaum*) auf die Menge der korrespondierenden Antwort-Nachrichten.

Da der Eingaberaum sehr hoch oder (bei Protokollen variabler Länge) nach oben unbe­schränkt sein kann, wird üblicherweise nur eine Teilmenge des Eingaberaums auf die Menge der korrespondierenden Antwort-Nachrichten abgebildet. Diese Abbildung wird im Folgenden als der *Fingerprint* eines Zielsystems bezeichnet.

Diese Definition bringt das Fingerprinting auf den Punkt: Es werden Anfrage-Nachrich­ten aus dem Eingaberaum an das Zielsystem versandt und die Antwort-Nachrichten empfangen. Die empfangenen Antwort-Nachrichten ergeben in Kombination mit den zuvor versandten Anfrage-Nachrichten den Fingerprint des Zielsystems und erlauben den Vergleich mit Fingerprints anderer Systeme, die auf den gleichen Anfrage-Nach­richten beruhen.

Bis zu diesem Punkt unterscheidet sich das Konzept noch nicht von bekannten Finger­printing-Ansätzen. Neu ist jedoch die Anwendung von Fuzzing zur Erzeugung von An­frage-Nachrichten. Dies hat folgende Gründe:

- Mittels Fuzzing kann die Erzeugung von Anfrage-Nachrichten automatisiert werden. Die menschliche Suche nach Implementierungsunterschieden etwa durch die Analyse von Changelogs oder durch Ausprobieren neuer Testfälle ent­fällt.
- Dabei werden auch bisher unentdeckte Eigenschaften im Antwort-Verhalten des Zielsystems ausgenutzt. Ein Fingerprinting auf Basis von Fuzzing ist somit po­tentiell genauer.

Jedoch ist ein Fingerprinting mit einer großen Zahl von Anfrage-Nachrichten nicht un-problematisch:

- Intrusion-Prevention-Systeme, die durch ungewöhnlichen Netzwerk-Traffic aktiv werden, könnten das Fingerprinting beeinträchtigen.[55]
- Durch ungünstige Anfrage-Nachrichten könnte das Zielsystem in seiner Verfügbarkeit beeinträchtigt werden (Denial of Service). Schließlich wird Fuzzing auch im Software-Testing zum Finden kritischer Schwachstellen eingesetzt.[56]
- Durch eine große Zahl von Anfrage-Nachrichten verlängert sich die Dauer des Fingerprintings. Da der Zeitraum für die Durchführung eines Penetrations-Tests begrenzt ist sollte die System- und Anwendungserkennung jedoch so schnell wie möglich abgeschlossen sein.[57]

Deswegen soll sich die Anwendung des Fingerprintings in zwei Schritte gliedern.

- Zunächst werden Fingerprints von bekannten Betriebssystemen und Diensten gesammelt.
- Für die Anwendung von Fingerprinting in Penetrations-Tests sollen nur jene Anfrage-Nachrichten verwendet werden, die einen Unterschied im Antwort-Verhalten verschiedener Implementierungen bewirken.

Das Vorgehen während des Fingerprintings wird im nächsten Kapitel beschrieben.

---

[55]   Vgl. Mörike, Michael (Hrsg.) (2004), S. 86
[56]   Vgl. Kapitel 2.5.1
[57]   Vgl. Ruef, Marc (2007), S. 59

## *3.2 Ein Phasenmodell für OS- und Service-Fingerprinting*

Für den gewählten Fingerprinting-Ansatz wurde der bereits beschriebene Fuzzing-Prozess[58] modifiziert. Der so entwickelte Fingerprinting-Prozess wird in Abbildung 3 dargestellt.

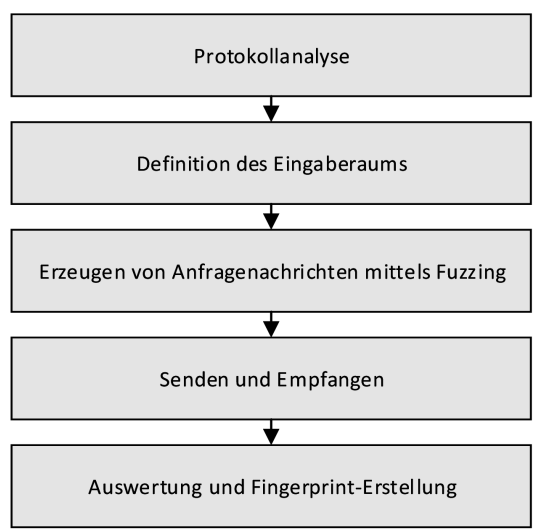

Abbildung 3: Der entwickelte Fingerprinting-Prozess

Die Teilschritte dieses Modells sollen zunächst allgemein besprochen werden. Im Anschluss werden die besonderen Ausprägungen der einzelnen Phasen für die Entwicklung einer eigenen OS- und Service-Fingerprinting-Software besprochen.

**Protokoll-Analyse**

Der erste Schritt des Fingerprinting-Prozesses ist die Analyse der beteiligten Protokolle und deren Einordnung in den jeweiligen Protokoll-Stack. Kommen verschiedene Protokolle oder nur eine Teilmenge eines Protokolls für das Fingerprinting in Betracht, so müssen die relevanten Protokolle bzw. Bestandteile eines Protokolls bestimmt werden. Die Protokoll-Analyse erfolgt stets manuell und kann nicht automatisiert werden.

---

[58]   Vgl. Kapitel 2.5.2

**Definition des Eingaberaums**

Der Erzeugung von Anfrage-Nachrichten geht die Definition des Eingaberaums voraus. In der Definition des Eingaberaums werden die Ergebnisse der Protokoll-Analyse auf den Punkt gebracht, indem die relevanten Protokollschichten und Protokollbestandteile genannt werden.

**Erzeugen von Anfrage-Nachrichten**

Für die Erzeugung der Anfrage-Nachrichten wird eine der bereits beschriebenen Fuzzing-Methoden[59] ausgewählt. Dabei wird für jeden Versuch die gleiche Eingabemenge vorausgesetzt, um die Ergebnisse des Fingerprintings reproduzierbar und untereinander vergleichbar zu halten.

**Senden und Empfangen**

Nun werden die erzeugten Anfrage-Nachrichten an das Zielsystem versandt, die zugehörigen Antwort-Nachrichten empfangen und der Fingerprint-Sammlung hinzugefügt.

**Auswertung des Antwort-Verhaltens**

Die Auswertung des Antwort-Verhaltens geschieht zu dem Zweck, Unterschiede in verschiedenen Implementierungen eines Protokolls zu finden, die eine Identifizierung ermöglichen. Dazu wird der Fingerprint des Zielsystems mit jedem der bereits erfassten Fingerprints verglichen. Diese Vergleiche setzen allerdings voraus, dass bereits Fingerprints erfasst wurden. Jeder dieser Vergleiche ergibt eine prozentuale Übereinstimmung zwischen dem neu erfasstem und dem bereits erfassten Fingerprint, die eine Zuordnung des Zielsystems zu einer konkreten Implementierung erleichtert.

Auf Grundlage des oben beschriebenen Phasenmodells werden in den folgenden Kapiteln konkrete Algorithmen, Methoden und Ansätze für den Einsatz im OS- und Service-Fingerprinting entwickelt.

---

[59] Vgl. Kapitel 2.5.3 - 2.5.6

## *3.3 Konzept für ein Fingerprinting von Betriebssystemen*

### 3.3.1 Protokoll-Analyse

Der erste Schritt des soeben beschriebenen Fingerprinting-Prozesses ist die Protokoll-Analyse. Für die Auswahl des Zielprotokolls des OS-Fingerprintings kommen verschiedene Schichten des TCP/IP-Stacks in Frage: Die Anwendungsschicht ist für ein OS-Fingerprinting eher ungeeignet. Zwar können im Rahmen eines Service-Fingerprintings auf der Anwendungsschicht unterschiedliche Verhaltensweisen eines Dienstes unter verschiedenen Betriebssystemen ausgemacht werden, doch kann in einer OS-Fingerprinting-Software nicht davon ausgegangen werden, dass ein bestimmter Dienst auf dem Zielsystem installiert ist.

Als ein geeigneteres Ziel des OS-Fingerprintings erscheinen vielmehr die Protokolle der Transportschicht und der Vermittlungsschicht, da sie in jedem TCP/IP-fähigen Host implementiert sind. Aufgrund dieser unterschiedlichen Anforderungen an ein Konzept für die Implementierung von Fingerprinting wurde ein dedizierter Ansatz für sowohl OS-Fingerprinting als auch Service-Fingerprinting verfolgt.

Als Zielprotokoll des Fuzzings entschied sich der Verfasser für TCP.[60] Der Vorteil von TCP ist, dass eine logische Ende-zu-Ende-Verbindung zwischen den kommunizierenden Systemen eingerichtet wird. Das bedeutet, dass der TCP-Header erst auf dem Zielsystem analysiert und etwaige Router auf dem Weg der Nachricht (vom Einsatz von Paketfiltern abgesehen) keinen Einfluss auf das Antwort-Verhalten haben.

Des Weiteren entschied der Verfasser, die TCP-Optionen in die Betrachtung mit einzubeziehen. Die Implementierung von TCP-Optionen ist für Betriebssysteme nicht obligatorisch. Die Reihenfolge, in der die implementierten Optionen im TCP-Header auftreten, ist zudem nicht definiert. Damit sind die TCP-Optionen ein erfolgversprechendes Ziel für Fingerprinting-Zwecke.

---

[60]   Vgl. Kapitel 2.2.3

Zuvor hatte der Verfassers versucht, Nachrichten mit unterschiedlichen Konfigurationen des IP-Headers zu versenden. Auf bestimmte Header-Werte wie eine ungewöhnliche IP-Versionsnummer oder IP-Header-Länge antworteten die Zielsysteme unterschiedlich mit ICMP-Nachrichten, die erklärten, dass die die IP-Nachricht vom Zielsystem abgewiesen wurde.

Dabei ergab sich jedoch folgendes Problem: Steht das Zielsystem in einem entfernten IP-Netzwerk, so durchläuft die Nachricht auf ihrem Weg mindestens einen Router.[61] Jeder Router entpackt die Nachricht und leitet die Informationen bis zur Vermittlungsschicht hoch. Dort wird die Nachricht analysiert und die beste Route zum Ziel ausgewählt. Bemerkt der Router eine Unregelmäßigkeit im IP-Header, so verwirft er das Paket und sendet ggf. selbst eine ICMP-Nachricht an den Sender zurück. Somit würde der Ansatz, den IP-Header in das Fuzzing mit einzubeziehen, nur das Verhalten des jeweiligen Next-Hop-Routers testen und musste deshalb verworfen werden.

### 3.3.2 Definition des Eingaberaums

Eingaberaum des OS-Fingerprintings sind alle möglichen Konfigurationen eines TCP-Segments unter Berücksichtigung der TCP-Optionen.

### 3.3.3 Erzeugen von Anfrage-Nachrichten

Für das OS-Fingerprinting soll der Ansatz des Generation-Based-Fuzzing verfolgt werden, das bedeutet, dass Testfälle auf Basis der Protokollspezifikationen erzeugt werden. Um die Geschwindigkeit des Fingerprintings zu verbessern, wird jedoch lediglich eine Auswahl (Range) des Eingaberaums generiert wie auf den nächsten Seiten genauer erklärt.

Um eine Vergleichbarkeit zwischen Antworten verschiedener Zielsysteme herzustellen, werden die Anfrage-Nachrichten genau einmal erzeugt, gespeichert und somit für jedes Fingerprinting wiederverwendet. Um Fingerprints verschiedener Mengen von Anfrage-Nachrichten miteinander zu vergleichen, müsste ein Versionsstempel verwendet werden, der die jeweilige Menge von Anfrage-Nachrichten kennzeichnet. Auf eine derartige

---

[61]   Vgl. Kapitel 2.2.1

Kennzeichnung wurde jedoch in der prototypischen Implementierung dieses Konzepts verzichtet.

Für die Erzeugung von Antwort-Nachrichten gilt es zunächst zu bestimmen, welche TCP-Felder mit automatisch generiertem Inhalt gefüllt werden sollen. In Frage kommen prinzipiell alle TCP-Felder. Jedoch muss das Zielport-Feld den Wert eines freien Ports auf dem Zielsystem annehmen. Damit das Zielsystem auf die Anfrage-Nachricht antwortet, muss außerdem das SYN-Flag im TCP-Header gesetzt sein.

Der Algorithmus zur Erzeugung der Anfrage-Nachrichten funktioniert folgendermaßen:

Gegeben sei eine Liste *fields* von TCP-Header-Feldern, die für die automatische Generierung ausgewählt wurden und eine numerische Variable s*tep.*

Rückgabewert sei eine Liste *F*, die für jedes ausgewählte TCP-Header-Feld eine Liste der schließlich in den Anfrage-Nachrichten verwendeten Feldinhalte enthält.

Für jedes Element dieser Liste wird eine neue Liste der verwendeten Feldinhalte generiert. Dabei wird folgendermaßen vorgegangen. Die Funktion *range* erzeugt für ein TCP-Feld systematisch Dezimalwerte aus dem Wertebereichs des Feldes. Übergeben wird der Funktion *range* der Start-Wert (sinnvollerweise stets 0, da im Wertebereich vorne angefangen wird) und der Ende-Wert (hier: die Größe des Feldes - 1, also das Ende des Wertebereichs, unter der Prämisse, dass ein Index mit 0 beginnt) der Auswahl.

Über die Angabe der Variable *step* (hier abgekürzt: s) wird die Menge der erzeugten Werte begrenzt, indem jeweils nur der s-te Wert aus dem Wertebereich generiert wird. Beträgt die Variable *step* etwa. 16, so enthält die generierte Menge die Dezimalwerte 0, 16, 32 ... bis zur maximalen Feldgröße.

Jede dieser Listen wird der Liste F hinzugefügt, aus der dann die Anfrage-Nachrichten erzeugt werden.

Dieser Erzeuger-Algorithmus kann auch durch folgenden Pseudocode beschrieben werden:

```
generate-fuzz (fields[], step):

F := new-list ()
n_fields := length-of ( fields )

for field_no := 0 to n_fields

     fuzz := range (0, fields[field_no] – 1, step)
     add-item-to-list (fuzz, F)

end-for

return F
```

Mit den generierten Listen von TCP-Feld-Inhalten  lassen sich nun Anfrage-Nachrichten erzeugen.

Sei $F$ die Liste der erzeugten TCP-Feld-Listen

mit $F_k$ als der k-ten TCP-Feld-Liste

und sei: $n = |F|$

so ist das kartesische Produkt:

$$\underset{k=1}{\overset{n}{\times}} F_k$$

die Menge der Anfrage-Nachrichten.

### 3.3.4 Senden und Empfangen

Es werden abwechselnd Anfrage-Nachrichten versendet und Antwort-Nachrichten empfangen, um eine Verkettung zwischen Anfragen und Antwort-Nachrichten herzustellen.

Dabei kommt ein weiterer Aspekt zum Tragen: Nachrichten, die per TCP oder UDP versandt werden, benötigen zwingend die Angabe eines Zielports. Somit ist ein Port-Scan vor dem eigentlichen Fingerprinting obligatorisch. Als Basis für das Port-Scanning wurde vom Verfasser das Half-Open-Scan-Verfahren[62] gewählt. Logischerweise funktioniert dieser Ansatz nur, wenn mindestens ein TCP-Port auf dem Zielsystem geöffnet ist. Andernfalls muss das Fingerprinting abbrechen.

### 3.3.5 Auswertung des Antwort-Verhaltens

Nach dem Empfang der Antwort-Nachrichten erfolgt die Auswertung des Antwort-Verhaltens, bei der der neu erstellte Fingerprint mit den vorhandenen Fingerprints verglichen wird.

Ein Fingerprint besteht aus vielen Antwort-Nachrichten, die weiter in konkrete Felder der TCP/IP-Header zerlegt werden. Für den Vergleich zweier Fingerprints sind nicht nur Felder des TCP-Headers von Bedeutung. Auch die Konfiguration des IP-Headers in den Antwort-Nachrichten als Reaktion auf die erzeugten Anfrage-Nachrichten birgt wertvolle Informationen, wie auch auf den nächsten Seiten gezeigt wird.

Der Vergleich zweier Antwort-Nachrichten ergibt die prozentuale Übereinstimmung der berücksichtigten Felder. Das arithmetische Mittel der Übereinstimmungen in den Antwort-Nachrichten ergibt dann die prozentuale Übereinstimmung zweier Fingerprints.

Der folgende Pseudocode soll den beschriebenen Ansatz verdeutlichen:

```
match_file (filename-current, filename-other, n_responses):
current := open-file (filename-current)
other   := open-file (filename-other)
total_match := 0

for response_no := 0 to n_responses
      match := match_response (response_no, other, current)
      total_match := total_match + match
end-for

total_match := 100 * total_match / n_responses

close-file (other)
close-file (current)

return total_match
```

---

[62]  Vgl. Kapitel 2.3.3

Dabei bedarf die hier verwendete Funktion *match_response* einer näheren Betrachtung. Der Vergleich von Feldern zweier Antwort-Nachrichten geschieht durch unterschiedliche Methoden. Die Auswahl der geeigneten Vergleichsmethode ist abhängig von der Art des Feldes. Dabei sind mindestens die folgenden Feldarten zu unterscheiden:

- Felder mit konstantem Inhalt
  Ist der Inhalt eines Feldes als Reaktion auf eine bestimmte Anfrage-Nachricht reproduzierbar, so genügt ein direkter Vergleich beider Felder, um deren Gleichwertigkeit zu bestimmten.

- Felder, deren Inhalt sich um einen bestimmten Wert häuft
  So ist etwa die TTL[63] eines IP-Paketes nicht in jeden Testaufbau reproduzierbar und hängt von der Position von Quell- und Zielsystem innerhalb der Netztopologie ab. Setzt man etwa eine Abweichung von +- 30 Hops voraus, so ergibt sich eine Häufung des TTL-Wertes auf bestimmte Wertebereiche.

- Felder, deren Inhalt als Teil einer Sequenz berechnet wird
  Beispiel hierfür ist die IP-Id, die Fragmente von IP-Paketen eindeutig identifiziert. Diese Anforderung wird teils als simpler Counter implementiert, teils aber auch durch den Einsatz von Pseudozufallszahlen erfüllt. Durch die unterschiedlichen Implementierungen können Hosts unterschiedlicher Betriebssysteme voneinander abgegrenzt werden.[64]

  Ein anderes Beispiel für eine Sequenz betrifft die Berechnung der TCP-ISN (Initial Sequence Number). Abermals werden fortlaufende Counter oder Pseudozufallszahlen für die Berechnung der ISN verwendet, teils wurde auch die Chiffrierung eines fortlaufenden Counters mit einem geheimen Schlüssel als Quelle für die ISN vorgeschlagen. [65]

  Als letztes Beispiel für Sequenzen sei die TCP-Timestamp-Option genannt. Auch der TCP-Timestamp wird in festgelegten Schritten erhöht, die sich jedoch in verschiedenen Betriebssystemen unterscheiden.[66]

---

[63]   Vgl. Kapitel 2.2.2
[64]   Vgl. Bellovin, Steven M. (2002), S. 272
[65]   Vgl. Bellovin, Steven M. (1989), S. 35
[66]   Vgl. Bursztein, E. (2007), S. 4

Um die benötigte Entwicklungszeit einzuschränken wurden in der Implementierung jedoch nur die Felder mit konstantem Inhalt berücksichtigt.

Welche Schlüsse die prozentuale Übereinstimmung zweier Fingerprints erlaubt wird in Kapitel 5 erörtert. Im nächsten Kapitel soll analog zur Vorgehensweise des OS-Fingerprintings ein Konzept für das Service-Fingerprinting erstellt werden.

## *3.4 Konzept für ein Fingerprinting von FTP-Servern*

### 3.4.1 Protokoll-Analyse

Ziel eines Service-Fingerprintings sind grundsätzlich alle Protokolle der Anwendungs­schicht. Dieses Konzept soll sich jedoch auf das File Transfer Protocol (FTP) beschrän­ken, da mit FTPMap 0.4 bereits eine Referenzimplementierung für das Fingerprinting von FTP-Servern existiert.

So kann versucht werden, die Zuverlässigkeit von FTPMap durch automatisch erzeugte Anfrage-Nachrichten zu übertreffen. Da FTPMap die Fingerprints von FTP-Servern hart kodiert, soll jedoch nicht etwa FTPMap erweitert werden, sondern eine eigene Fin­gerprinting-Software entwickelt werden, die eine leichte Erweiterbarkeit um Finger­prints neuer FTP-Server bietet.

Eine FTP-Sitzung besteht aus zwei getrennten TCP-Verbindungen, einer Kontrollver­bindung für die Steuerung der Anwendung und einer Datenverbindung, über die die an­geforderten Dateien transferiert werden. Ziel dieser Betrachtung ist ausschließlich die Kontrollverbindung des FTP-Servers, nicht jedoch die Datenverbindung, da die vom FTP-Server bereitgestellten Dateien für das Fingerprinting irrelevant sind.

### 3.4.2 Definition des Eingaberaums

> Eingaberaum sind jegliche Nachrichten, die über die FTP-Kontrollverbindung ausgetauscht werden.

### 3.4.3 Erzeugen von Anfrage-Nachrichten

Es wird der Ansatz des Mutation-Based-Fuzzings verfolgt. Dabei werden zunächst gül­tige Befehle typischer FTP-Server identifiziert. Dies umfasst insbesondere auch FTP-Befehle, die nicht standardisiert sind und daher ein gutes Unterscheidungsmerkmal bil­den. Dabei sollen nur solche Befehle ausgewählt werden, die die Kontrollverbindung betreffen. Kommandos, die Dateien herunterladen, hochladen oder Verzeichnisse erstel­len, sollen nicht Ziel der Betrachtung sein, da das Antwort-Verhalten von den

Dateistrukturen des Zielsystems abhinge und sogar durch den Fingerprinting-Vorgang beeinflusst werden könnte.

Ein Fuzzer-Algorithmus erzeugt nun für jeden ausgesuchten FTP-Befehl und jede mögliche Länge bis zu einer vom Anwender definierten oberen Schranke ein Argument. So wird nicht nur ein Test auf unterschiedliche Eingabemengen, sondern auch ein Test auf das Verhalten bei unterschiedlich langen Argumenten, wie auch in fortgeschrittenen Service-Fingerprinting-Tools benutzt[67], durchgeführt.

Von jeder erzeugten Anfrage-Nachricht für einen gegebenen FTP-Befehl und eine gegebene Länge werden *n* Instanzen erzeugt, die jeweils *m*-mal mutieren. Beide Parameter können durch den Anwender der Software bestimmt werden. Durch die mehrfache Mutation wird die konkrete Eingabemenge des Fingerprintings vergrößert und damit die Genauigkeit des Fingerprintings potentiell verbessert.

Zur Mutation wird ein Algorithmus verwendet, der die übergebene Nachricht bei jedem Aufruf entweder um ein zufällig gewähltes Zeichen verlängert, zufällig ein Zeichen ändert oder zufällig ein Zeichen aus der Nachricht entfernt.

Die Anfrage-Nachrichten werden grundsätzlich nur einmal für jede Versuchsreihe erzeugt. So kann jedes Zielsystem mit der gleichen Eingabemenge getestet werden, um die Vergleichbarkeit der Antwort-Nachrichten zu gewährleisten.

### 3.4.4 Senden und Empfangen

Das Versenden der Anfrage-Nachrichten geschieht wie beim OS-Fingerprinting durch ein synchronisiertes Anfrage-Antwort-Verhalten. Das bedeutet, dass unmittelbar nach dem Versand einer Anfrage-Nachricht die zugehörige Antwort-Nachricht empfangen und gespeichert wird, um eine Verkettung zwischen Anfrage- und Antwort-Nachrichten herzustellen.

Beim Start des Fingerprintings wird dem Programm IP-Adresse bzw. Hostname des Zielsystems und FTP-Server-Port des untersuchten Dienstes übergeben. Fehlt die Angabe eines Ports, wird der Standard-Port des FTP verwendet (TCP/21).

---

[67]   Vgl. Kapitel 2.3.5

### 3.4.5 Auswertung des Antwort-Verhaltens

Als Vergleichskriterium kommen bei FTP die Statuscodes und die zugehörigen Wordings (Statusbeschreibungen) in Frage, die die Felder der Antwort-Nachricht bilden. Um die Entwicklungszeit sinnvoll zu begrenzen werden die Wordings jedoch nicht in der Implementierung des Konzepts berücksichtigt.

Die Auswertung des Antwort-Verhaltens erfolgt über einen Algorithmus, der jeden Feldinhalt einer Fingerprint-Datei mit jedem Feldinhalt einer anderen Fingerprint-Datei vergleicht. Zu diesem Zweck wird der bereits in Kapitel 3.3.5 für das OS-Fingerprinting formulierte Algorithmus herangezogen, der im Ergebnis die prozentuale Übereinstimmung beider Fingerprint-Dateien errechnet und zurückliefert.

## *3.5 Anforderungen an eine Implementierung des Konzepts*

Bevor die konzipierte Software programmiert wird, sollen Anforderungen an eine Implementierung definiert werden. In Kapitel 5 sollen diese Anforderungen als Grundlage für eine Bewertung der programmierten Software dienen.

**Zuverlässigkeit**

Die Fingerprinting-Software soll Betriebssysteme und Dienste zuverlässig erkennen. Das bedeutet, dass Betriebssysteme und Dienste unterschiedlicher Produktreihen stets zu einem unterschiedlichen Fingerprint führen. Dabei sollte der Unterschied zwischen zwei verwandten Releases einer Produktreihe kleiner sein als der Unterschied zwischen zwei Betriebssystemen bzw. Diensten unterschiedlicher Produktreihen.

**Portabilität**

Die Fingerprinting-Software soll auf verschiedenen Betriebssystemen verwendet werden können.

**Erweiterbarkeit**

Die Fingerprinting-Software soll leicht um Fingerprints neuer Betriebssysteme und Dienste erweitert werden können.

**Bedienbarkeit**

Die Fingerprinting-Software soll leicht und intuitiv bedienbar sein.

Unter Berücksichtigung dieser Anforderungen kann mit der Implementierung des Konzeptes begonnen werden.

# 4 Implementierung der konzipierten Fingerprinting-Software

## 4.1 Komponenten und Struktur der Implementierung

Die Implementierung der OS- und Service-Fingerprinting-Software geschieht über die folgend abgebildete Struktur.

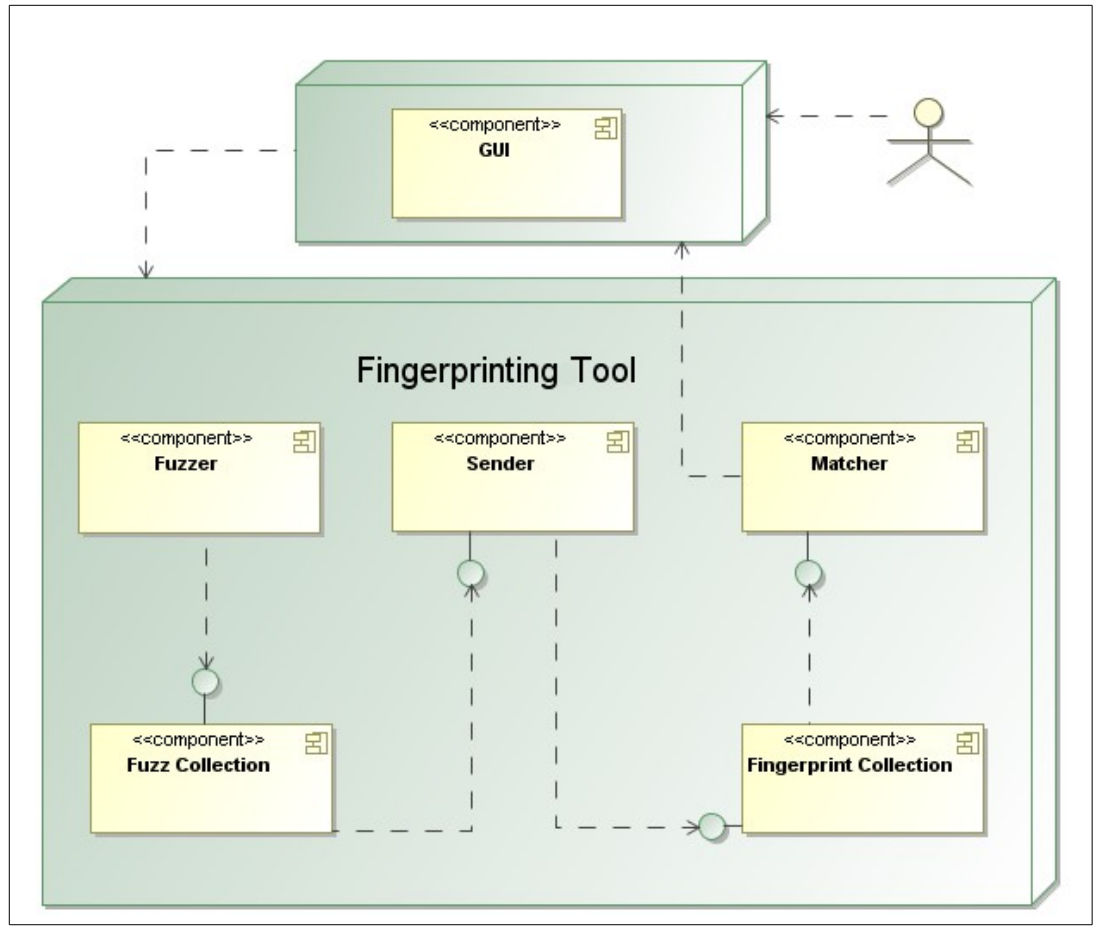

Abbildung 4: UML-Diagramm der Implementierung

Der Benutzer interagiert mit der GUI, konfiguriert die Eingabeparameter und startet das Fingerprinting.

Die *Fuzzer*-Komponente erzeugt die Anfrage-Nachrichten und speichert sie in der *Fuzz-Collection*.

Die Sender-Komponente liest die Anfrage-Nachrichten aus der Fuzz-Collection aus, versendet sie über das Netzwerk und sammelt die Antwort-Nachrichten in einer Fingerprint-Datei.

Die *Matcher*-Komponente vergleicht den neu erzeugten Fingerprint schließlich mit der vorhandenen *Fingerprint-Collection* und berechnet eine Übereinstimmung mit vorhandenen Betriebssystemen/Diensten.

Diese Information wird in der GUI angezeigt und ermöglicht dem Benutzer, Rückschlüsse auf das Betriebssystem bzw. einen Dienst des Zielsystems zu ziehen.

Schließlich wird der neue Fingerprint für spätere Vergleiche der Fingerprint-Collection übergeben. So wird den Anforderungen an die Erweiterbarkeit der Fingerprinting-Software[68] Rechnung getragen.

Die konkrete Implementierung dieser Struktur zum Zweck des OS-Fingerprintings wird im folgenden Kapitel beschrieben.

---

[68]   Vgl. Kapitel 3.5

## *4.2 Implementierung der OS-Fingerprinting-Software*

### 4.2.1 Auswahl der Entwicklungswerkzeuge

Für die Implementierung des OS-Fingerprintings wurde zunächst der Paket-Generator Hping begutachtet. Hping ermöglicht es, nach Vorgaben des Benutzers IP-Datagramme mit beliebiger Nutzlast zusammenzustellen. Ab Version 3 beinhaltet Hping auch einen Interpreter für die Skript-Sprache Tcl (Tool Command Language, ausgesprochen: tickel), der die Entwicklung von Netzwerkanalyse-Tools und Sicherheitssoftware weiter vereinfachen soll.[69]

Tcl[70] wurde von John Ousterhout als „general purpose"-Sprache, d. h. als Programmiersprache für den Einsatz in beliebigen Anwendungsfeldern entwickelt. Tcl ist quelloffen und bietet ein mächtiges Modul-Konzept. Auf diese Weise ist eine leichte Erweiterung der Fingerprinting-Software möglich. Tcl ist für Microsoft Windows (ab Windows 95), für MacOS (ab Version 8) und alle gängigen UNIX-artigen Betriebssysteme verfügbar. So können die Anforderungen an die Portabilität[71] erfüllt werden. Die Entwicklung grafischer Benutzeroberflächen wird durch die nahtlose Integration der Grafik-Bibliothek Tk in den Lieferumfang von Tcl begünstigt. Somit können auch die Anforderungen an die Bedienbarkeit[72] erfüllt werden.

Mit der Zusammenstellung der benötigten Entwicklungstools konnte die Arbeit am Quellcode begonnen werden. Das fertige Programm hat einen Umfang von ca. 250 LOC.

### 4.2.2 Das APD-Format

Durch das APD-Format (ARS Packet Description) können IP-Pakete menschenlesbar spezifiziert werden. Dabei interpretiert Hping eine Zeile im APD-Format als eine Nachricht, die verschiedene Schichten des TCP/IP-Stacks beschreibt.
Abstrakt kann die Syntax einer Schicht folgendermaßen dargestellt werden:

---

[69]   Vgl. Sanfilippo, Salvatore (2005)
[70]   Vgl. Jakobs, Holger (2004), S.2
[71]   Ebenda
[72]   Ebenda

| layer_type{field_1=value_1,field_2=value_2,...,field_n=value_n} |
|---|

Dabei werden unterschiedliche Protokoll-Schichten durch ein +-Symbol miteinander verbunden. Ein konkretes Beispiel für eine DNS-Nachricht auf Basis des APD-Formats sei im Folgenden dargestellt:

| ip{daddr=192.168.1.2}+udp{sport=53,dport=53}+data{file=./dns.packet} |
|---|

Fehlende Feldangaben, die wie die Prüfsumme eines Headers von Hping berechnet werden können, werden automatisch ergänzt.[73]

### 4.2.3 Der Fuzzer

Der Fuzzer-Komponente besteht im Wesentlichen aus den Funktionen *fuzz_tcp* und *fuzz_tcp_options*. Dabei wird zunächst für jedes Feld bzw. jede Option eine Liste möglicher Werte erzeugt. Die von der TCL-Community entwickelte Funktion *crossProduct* erzeugt dann das kartesisches Produkt dieser Werte in Form einer neuen Liste, die alle verwendeten Konfigurationen des TCP-Headers bzw. der TCP-Optionen enthält.

Die Funktion *generate_packets* erzeugt nun aus diesen Konfigurationen fertige IP-Datagramme mit Dummy-Werten für die IP-Adressfelder und den Zielport. Diese Felder sind von den am Fingerprinting beteiligten Systemen abhängig und werden erst beim Versand gesetzt. Die so präparierten Pakete werden unter Verwendung des APD-Formates in der Fuzz-Collection gespeichert.

### 4.2.4 Die Fuzz-Collection

Die Fuzz-Collection ist eine Textdatei, die die erzeugten Anfrage-Nachrichten im APD-Format beinhaltet. Jede Zeile entspricht genau einer Anfrage-Nachricht.

---

[73]  Vgl. Sanfilippo, Salvatore (2005)

**4.2.5 Der Sender**

Der Sender wird durch die Funktion *send_and_receive* repräsentiert. Nach jedem Versand eines Paketes wird die zugehörige Antwort-Nachricht empfangen. Durch die Verwendung von Raw-Sockets werden jedoch auch Pakete empfangen, die keine Antwort auf die versendete Anfrage-Nachricht sind. Daher müssen die Antwort-Nachrichten aus der Menge der empfangenen Pakete herausgefiltert werden. Dies geschieht durch eine Untersuchung des Quellports und der Quell-IP-Adresse der empfangenen Pakete. Geht nach einer bestimmten Zahl empfangener Pakete bzw. nach Verstreichen eines Timeouts keine Antwort-Nachricht ein, so wird angenommen, dass das Zielsystem nicht auf die versendete Anfrage-Nachricht antwortet. Dieser Timeout ist ggf. den physikalischen Gegebenheiten des Zielnetzwerks anzupassen: In einem LAN können in der Regel kürzere Timeouts als in einem WAN gewählt werden, um so die Performance des Fingerprintings zu verbessern.

**4.2.6 Die Fingerprint-Collection**

Die Fingerprint-Collection ist das Verzeichnis aller gesammelten Fingerprints. Jeder Fingerprint ist eine Textdatei im APD-Format, die in Zeile n die Antwort-Nachricht auf die n-te Anfrage-Nachricht enthält. Eine Leerzeile deutet dabei auf eine ausgebliebene Antwort des Zielsystems hin.

**4.2.7 Der Matcher**

Der Matcher wird durch die Funktion *match_os* implementiert, die den erzeugten Fingerprint mit allen Fingerprints der Fingerprint-Collection vergleicht. Beim Vergleich zweier Fingerprints werden die Antworten auf eine Anfrage-Nachricht feldweise verglichen. Dabei wurde die Erkennung von Häufungen und Sequenzen[74] in dieser Implementierung nicht berücksichtigt und ist etwaigen späteren Versionen vorbehalten. Im Ergebnis wird eine Liste der fünf Fingerprints mit den höchsten prozentualen Übereinstimmungen ausgegeben.

---

[74]  Vgl. Kapitel 3.3.5

## *4.3 Implementierung der Service-Fingerprinting-Software*

### 4.3.1 Auswahl der Entwicklungswerkzeuge

Um die Anforderungen an eine schnelle Entwicklungszeit zu erfüllen, entschied sich der Verfasser für die Verwendung der bestehenden Fuzz-Testing-API „Antiparser" als Grundlage eigener Entwicklungen. Antiparser ist eine API für die Programmiersprache Python und wurde 2005 von David McKinney veröffentlicht.[75]

Die Programmiersprache Python wurde zu Beginn der 1990er Jahre von Guido van Rossum entworfen. Ziel der Entwicklung war es, eine Sprache zu schaffen, die möglichst einfach und gleichzeitig mächtig sein sollte.[76] Aufgrund dieser Eigenschaften eignet sich Python hervorragend für den RAD-Ansatz.

Antiparser enthält die Referenz-Implementierung eines Fuzzers für FTP-Server, den sogenannten „Evil FTP Client". Als Basis für die Erzeugung von Antwort-Nachrichten dient dabei eine Menge gültiger FTP-Befehle, von denen der Fuzzer jeweils unterschiedliche Varianten erzeugt.

Mit diesem Mitteln konnte die Arbeit an einer eigenen Fingerprinting-Software begonnen werden. Das implementierte Programm hat einen Umfang von ca. 350 LOC.

### 4.3.2 Der Fuzzer

Der Fuzzer setzt sich aus zwei Modulen zusammen. Zum einen erzeugt die Antiparser-Bibliothek FTP-Befehle mit Argumenten unterschiedlicher Länge, zum anderen erzeugt die Funktion *mutate* verschiedene Instanzen der erzeugten Befehle einer bestimmten Länge. Diese Befehle werden in der Fuzz-Collection gespeichert und beim Fingerprinting eines Hosts wieder abgerufen.

---

[75]   Vgl. McKinney, David (2005)
[76]   Vgl. Kaiser und Ernesti (2007)

### 4.3.3 Die Fuzz-Collection

Die Fuzz-Collection ist eine Textdatei, die alle erzeugten Anfrage-Nachrichten enthält. Dabei wird in jeder Zeile der Datei genau eine Anfrage-Nachricht gespeichert.

### 4.3.4 Der Sender

Der Sender baut eine TCP-Verbindung zum Zielsystem auf und loggt sich anonym auf dem FTP-Server ein. Ist kein anonymer Zugriff möglich, muss das Fingerprinting abbrechen. Anschließend werden in einer Schleife abwechselnd Anfrage-Nachrichten aus der Fuzz-Collection versandt, die zugehörige Antwort-Nachricht empfangen und in der Fingerprint-Collection gespeichert.

### 4.3.5 Die Fingerprint-Collection

Zum gegenwärtigen Zeitpunkt speichert die implementierte Service-Fingerprinting-Software lediglich den FTP-Statuscode einer Antwort-Nachricht.

Die Speicherung der Statuscodes geschieht in Textdateien der Form:

```
Statuscode-1
Statuscode-2
...
Statuscode-N
```

für N Antwort-Nachrichten.

### 4.3.6 Der Matcher

Der Matcher iteriert über die Zeilen zweier Fingerprint-Dateien und ermittelt die Übereinstimmung der empfangenen Statuscodes. Wie beim OS-Fingerprinting werden dem Anwender die fünf Fingerprints mit dem höchsten Übereinstimmungsgrad gezeigt.

## *4.4 Implementierung der grafischen Benutzeroberfläche*

Auf Grundlage der Grafik-Bibliothek Tk wurde eine eigene GUI entwickelt. Dabei galt es zunächst, die nötigen Informationen für die Eingabemaske zu bestimmen. In der Standardansicht sind nur die Eingabeparameter der Sender-Komponente sichtbar.

Für das OS-Fingerprinting sind dabei die folgenden Eingabefelder notwendig.

**Host**
Über den Hostnamen bzw. eine IP-Adresse wird das Zielsystem adressiert.

**Betriebssystem (falls bekannt)**
Ist das Betriebssystem des Zielsystems bereits bekannt, so sollte eine möglichst genaue Angabe in diesem Feld gemacht werden. Auf diese Weise können neue Fingerprints aufgenommen werden, die die Erkennungs-Wahrscheinlichkeit des Fingerprintings verbessern.

Für das Service-Fingerprinting dagegen sind folgende Eingabefelder relevant.

**Host**
Wiederum ist die Eingabe des Hostnamens nötig, um das Zielsystem zu adressieren.

**Port**
Auch wenn FTP standardmäßig Port 21 für die Kontroll-Verbindung nutzt, soll der Port des Zielsystems konfigurierbar sein.

**Release (falls bekannt)**
Wie beim OS-Fingerprinting kann dem Programm ein bekanntes FTP-Server-Release übergeben werden, um die Fingerprint-Sammlung zu erweitern.

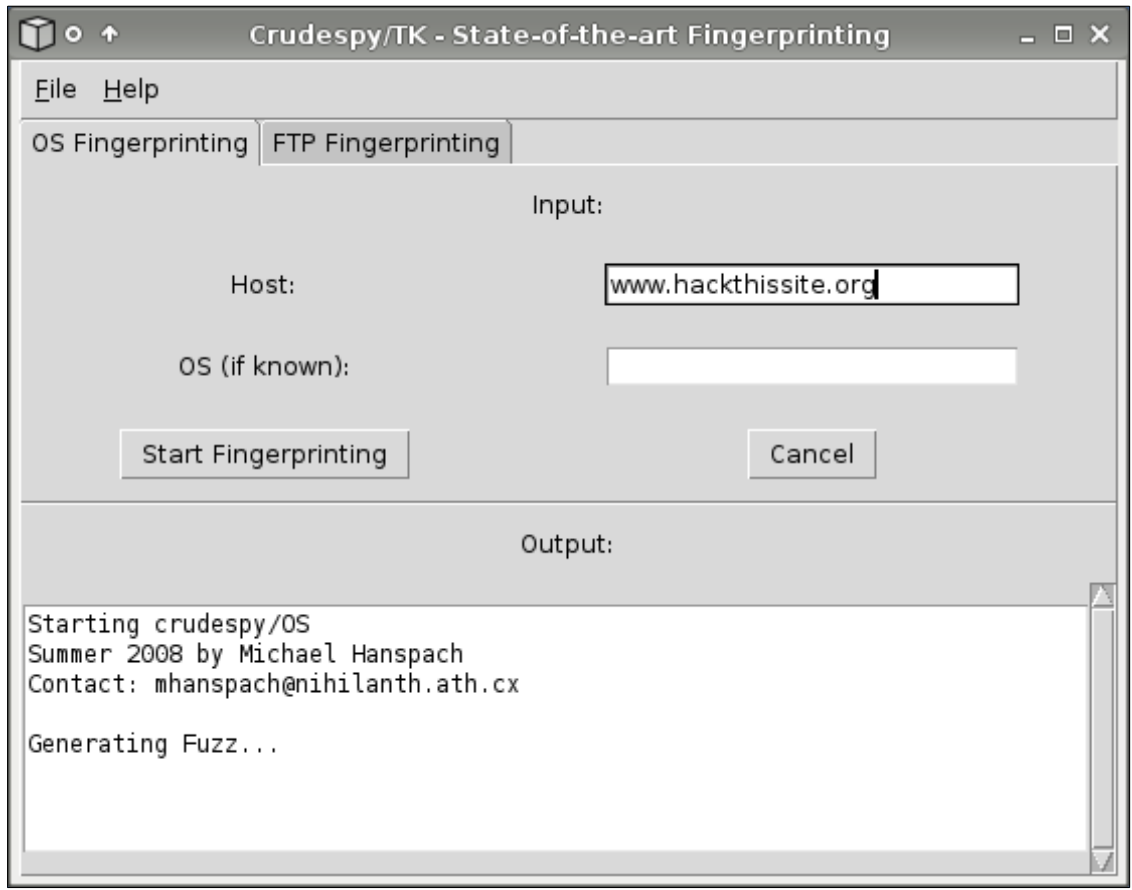

Abbildung 5: Screenshot der entwickelten grafischen Benutzeroberfläche

Die Ausgabe des Programms soll in einem Textfeld erfolgen. Mit diesen Angaben konnte die grafische Benutzeroberfläche auf Basis von Tcl/Tk implementiert werden.

Mit der konkreten Implementierung kann die Fingerprinting-Software getestet werden, um die Zuverlässigkeit des konzipierten Fingerprinting-Ansatzes zu bewerten.

# 5 Test und Bewertung der implementierten Fingerprinting-Software

## 5.1 Test der Zuverlässigkeit des OS-Fingerprintings

Für den Testaufbau des OS-Fingerprintings wurden die Betriebssysteme Debian GNU/Linux 4.0, Kernel 2.6.18, MirOS 80316, OpenSolaris 2008.5, Windows XP Professional SP2 und DragonFly BSD 2.0 installiert. Da die implementierte Fingerprinting-Software einen offenen TCP-Port voraussetzt, wurde für jedes getestete Betriebssystem der Webserver Apache 2.2.9 installiert.

Die Ergebnisse des Fingerprintings werden in Tabelle 3 dargestellt.

Tabelle 3: Verhaltens-Übereinstimmungs-Matrix für Betriebssysteme

| OS | Debian | MirOS | OpenSolaris | Windows XP | DragonFly BSD |
|---|---|---|---|---|---|
| Debian | 100,00% | 75,55% | 73,30% | 51,62% | 86,01% |
| MirOS | 75,55% | 100,00% | 59,21% | 64,33% | 73,73% |
| OpenSolaris | 73,30% | 59,21% | 100,00% | 47,99% | 71,49% |
| Windows XP | 51,62% | 64,33% | 47,99% | 100,00% | 57,50% |
| DragonFly BSD | 86,01% | 73,73% | 71,49% | 57,50% | 100,00% |

Dieses Ergebnis bleibt auch nach mehreren Testläufen reproduzierbar. Anhand dieser Tabelle wird sichtbar, dass sich TCP/IP-Stacks verschiedener Betriebssysteme deutlich in ihrem Antwort-Verhalten unterscheiden und mittels der durch Fuzzing generierten Anfrage-Nachrichten sicher voneinander abgegrenzt werden können.

Da Betriebssysteme mit teils sehr unterschiedlicher Historie ausgewählt wurden, bleibt die Frage offen, ob auch unterschiedliche Revisionen eines spezifischen Betriebssystems voneinander abgegrenzt werden können.

Zu diesem Zweck wurden innerhalb des installierten Betriebssystems Debian GNU/Linux 4.0 verschiedene Releases des Kernels installiert. Zunächst wurde dabei auf die offiziell gepatchten Kernel-Images 2.6.18-6-486 und 2.6.24-etchnhalf.1-686 der Debian-Distribution zurückgegriffen. Im Test ergab sich zwischen beiden Kernels eine Übereinstimmung von 95,94%. Diese Zahl ermöglicht zwei Schlüsse:

1. Es gibt Kernel-Releases, die sich definitiv in ihrem TCP/IP-Stack voneinander unterscheiden und mit der hier entwickelten Fingerprinting-Software voneinander unterschieden werden können.

2. Dieser Unterschied ist gering genug, um auch bei unterschiedlichen Kernel-Releases eine passende Zuordnung zu dem verwendeten Betriebssystem zu ermöglichen.

Jedoch sei angemerkt, dass diese Fähigkeit zur Unterscheidung von Kernel-Releases begrenzt ist. So konnte zwischen den offiziellen Linux-Versionen 2.6.21.7, 2.6.22.19, 2.6.23.17 und 2.6.24.7 der Linux Kernel Archives (www.kernel.org) kein Unterschied ausgemacht werden. Möglicherweise könnten diese Releases durch eine Erweiterung des Matchers um neue Metriken unterschieden werden. Es ist jedoch auch möglich, dass zwischen diesen Kernel-Releases keine relevanten Unterschiede im TCP/IP-Stack existieren.

Ferner ist interessant, ob es mit der entwickelten Software gelingt, die Genauigkeit des bekannten Port-Scanners Nmap beim OS-Fingerprinting[77] zu übertreffen. Zu diesem Zweck musste zunächst ein Testfall gefunden werden, der von Nmap nicht unterschieden werden kann: Die beiden Kernel-Releases 2.6.25-gentoo-r7 und 2.6.26-rc5-mm3 der Gentoo-Linux-Distribution verhielten sich unter Nmap 4.68 gleich. Beim Testlauf mit der hier entwickelten Fingerprinting-Software ergab sich jedoch ein reproduzierbarer Unterschied von 0,02% in den Fingerprints beider Betriebssysteme. Dies mag gering wirken, doch das wird durch die große Zahl versendeter Anfrage-Nachrichten relativiert. Somit konnte mindestens ein signifikanter Testfall zur Unterscheidung zweier Kernel-Releases neu erschlossen werden.

Nachdem die Zuverlässigkeit des OS-Fingerprinting-Ansatzes dargelegt wurde, soll ebenfalls die Zuverlässigkeit der Service-Fingerprinting-Software gezeigt werden.

---

[77]  Vgl. Kapitel 2.3.3

## *5.2 Test der Zuverlässigkeit des Service-Fingerprintings*

Mit Hilfe der entwickelten Service-Fingerprinting-Software sollen Unterschiede im Verhalten verschiedener FTP-Server gesucht werden. Für diesen Zweck wurden die FTP-Server GLFTPD 2.0.1, Net-FTPServer 1.122, Netkit-FTPD 0.17, PROFTPD 1.3.1, Pure-FTPD-1.0.21 und VSFTPD 2.0.5 auf dem Zielsystem installiert.

Zunächst werden Fingerprints für alle zu testende FTP-Server gesammelt. Zu diesem Zweck wird noch die gesamte Fuzz-Menge vom Sender verarbeitet. Im Produktivbetrieb dagegen sollen nur jene Testfälle eingesetzt werden, die nachweislich als Unterscheidungsmerkmal dienen. Weiterhin wird der Fingerprinting-Prozess für jeden zu testenden FTP-Server ein weiteres Mal durchlaufen. Die entwickelte Software gibt daraufhin bei jedem Durchlauf eine Liste der fünf in ihrem Verhalten ähnlichsten Server aus. Die Ergebnisse werden in Tabelle 4 übertragen.

Tabelle 4: Verhaltens-Übereinstimmungs-Matrix für FTP-Server

| **FTP-Server** | GLFTPD | Net-FTPServer | Netkit-FTPD | PROFTPD | Pure-FTPD | VSFTPD |
|---|---|---|---|---|---|---|
| GLFTPD | 100,00 % | 37,82% | 81,41% | 51,54% | 37,92% | 46,15% |
| Net-FTPServer | 37,82% | 100,00% | 38,46% | 56,41% | 74,36% | 58,97% |
| Netkit-FTPD | 81,41% | 38,46% | 100,00% | 51,92% | 43,59% | 39,10% |
| PROFTPD | 51,54% | 56,41% | 51,92% | 100,00% | 51,28% | 51,32% |
| Pure-FTPD | 37,92% | 74,36% | 43,59% | 51,28% | 100,00% | 66,67% |
| VSFTPD | 46,15% | 58,97% | 39,10% | 51,32% | 66,67% | 100,00% |

Auch nach mehreren Durchläufen bleibt dieses Ergebnis konstant. Diese Zahlen erlauben die folgenden Rückschlüsse:

- Das Antwort-Verhalten der verschiedenen FTP-Server ist signifikant unterschiedlich und erlaubt tatsächlich ein Fingerprinting mit den durch Fuzzing erzeugten Daten.

- Stimmen die Fingerprints zweier Server zu 100% überein, so kann man davon ausgehen, dass es sich um das gleiche Produkt handelt.

Was ist jedoch, wenn sich die Fingerprints zweier Zielsysteme unterscheidet? Prinzipiell sind drei Fälle denkbar:

- Die Zielsysteme verwenden unterschiedliche FTP-Server. Die prozentuale Übereinstimmung der Fingerprints ist ein guter Indikator dafür, ob es sich um zwei völlig verschiedene Dienste handelt oder nicht. Fraglich ist jedoch, wo die Grenze gezogen werden soll. Wie aus Tab. 4 ersichtlich, beträgt der höchste Übereinstimmungsgrad zwischen zwei verschiedenen Servern 81,41%. Auf Grundlage oben genannter Daten könnte etwa eine Grenze bei 90% gezogen werden. Um zu einem hinreichend genauen Ergebnis zu kommen, müssten jedoch wesentlich mehr FTP-Server getestet werden.
- Die Zielsysteme verwenden unterschiedliche Betriebssysteme. Dies kann einen erheblichen Einfluss auf das Ergebnis haben. In einem Vergleich des Pure-FTPD 1.0.21 unter Linux und Windows (innerhalb der weit bekannten Cygwin-Umgebung) ergab sich lediglich eine Übereinstimmung von 72,91% beider Fingerprints. Somit scheint es sinnvoll, Fingerprints für beide Server-Varianten vorzuhalten.
- Die Zielsysteme verwenden unterschiedliche Releases der gleichen Software. So ergibt sich zwischen den Versionen 2.0.5 und 2.0.7 des VSFTPD eine reproduzierbare Verhaltens-Übereinstimmung von 97,80%. Der Unterschied von 2,2% mag gering wirken, ist jedoch ein sicheres Unterscheidungsmerkmal. Die Möglichkeit, einzelne Releases voneinander zu unterscheiden, ist dennoch begrenzt, da manche Versionssprünge schlicht keine Änderungen im Antwort-Verhalten beinhalten. So konnte mit der konzipierten Lösung kein Verhaltens-Unterschied zwischen den Versionen 1.0.21 und 1.0.11 des Pure-FTPD ausgemacht werden.

Schließlich stellt sich die Frage: Ist das entwickelte Fingerprinting-Tool tatsächlich zuverlässiger als die bestehenden Service-Fingerprinting-Tools wie THC Amap und FTPmap? Dieser Frage soll im Folgenden nachgegangen werden.

Vor dem Testen wurde die Banner-Ausgabe jedes FTP-Servers deaktiviert, um die Vorteile eines Fingerprintings gegenüber der Anwendung von Banner-Grabbing zu demonstrieren. Über Banner-Grabbing lassen zwar sich schnell präzise Ergebnisse erreichen, jedoch ist die Deaktivierung des Banners gängige Praxis im Internet und fast jeder erhältliche FTP-Server bietet eine entsprechende Einstellung an. Die einzige Ausnahme in diesem Test war der Netkit-FTPD. Um die Ausgabe des Banner-Strings zu deaktivieren, musste der Quellcode angepasst und die Server-Software neu kompiliert werden.

Die Fingerprinting-Software THC Amap 5.2 ist ohne Ausgabe von Banner-Strings nutzlos. Trotz der Verwendung von im Vorhinein gespeicherter Erkennungsmerkmale konnte nach der Deaktivierung der Banner-Strings keiner der getesteten FTP-Server von Amap 5.2 identifiziert werden. Als einzige Information kann die Verwendung des File Transfer Protocol der Ausgabe entnommen werden. Dagegen konnte mit der selbst entwickelten Fingerprinting-Software jede der getesteten Server-Varianten reproduzierbar identifiziert und durch eine prozentuale Übereinstimmung klar von anderen FTP-Servern abgegrenzt werden.

Das Programm FTPmap 0.4 ist eine dedizierte FTP-Fingerprinting-Software: Es werden Standard-Testfälle aus einer Konfigurationsdatei gelesen, an den entfernten FTP-Server versandt und die Antworten ausgewertet. Nachdem Fingerprints für alle eingesetzten FTP-Server erstellt wurden, konnte FTPmap 0.4 jeden Dienst bei späteren Tests wiedererkennen. Um eine spezifische Server-Software zu erkennen, ist dieses Vorgehen somit völlig ausreichend. Fraglich ist jedoch, ob beide Ansätze gleich gut darin sind, unterschiedliche Releases zu identifizieren.

Gelingt es, innerhalb der Testreihen zwei Releases zu finden, die von FTPmap 0.4 nicht wirkungsvoll unterschieden werden können, jedoch durch den Fuzzing-Ansatz identifiziert werden können, so ist eine Verbesserung von Service-Fingerprinting durch systematisch erzeugte Anfrage-Nachrichten möglich.

Die Vergleichstests umfassten die Installation verschiedener FTP-Server-Releases in der gleichen Systemumgebung, den Vergleich zwischen diesen Releases mit FTPmap 0.4 und der hier entwickelten Fingerprinting-Lösung sowie den Vergleich der Wirksamkeit beider Tools. Dabei gelang es tatsächlich, im Vergleich zweier FTP-Server die Zuver-lässigkeit des Service-Fingerprintings zu verbessern. Wie bereits erwähnt, ist die entwickelte Fingerprinting-Software die Versionen 2.0.5 und 2.0.7 von VSFTPD unter Linux 2.6.15 (Gentoo) zu identifizieren. Der Verhaltensunterschied beträgt dabei reproduzierbar 2,2%. Diese Unterscheidung ist mit FTPmap schon nicht mehr möglich.

Zu FTPmap 0.4 gehört eine Datei mit Anfrage-Nachrichten, die an den FTP-Server des Zielsystems versandt werden. Die Schwäche des Programms liegt letztendlich in diesen manuell gewählten Anfrage-Nachrichten. Somit ist es auch möglich, FTPmap durch Einsatz der mittels Fuzzing generierten Anfragen zu verbessern und den Vorsprung der hier entwickelten Fingerprinting-Software wieder aufzuholen. Dafür extrahiert die vom Verfasser geschaffene Fingerprinting-Lösung genau jene Testfälle, deren Antwort-Nachrichten sich in verschiedenen Servern und Server-Releases unterscheiden und somit ein Identifizierungsmerkmal bieten.

Das selbe Vorgehen ist auch beim HTTP-Fingerprinting möglich. Hmap ist eine spezialisierte HTTP-Fingerprinting-Software. Genau wie bei FTPmap werden manuell erzeugte Testfälle verwendet, was die Vermutung nahe liegen lässt, dass es ebenfalls Verbesserungspotentiale in der Erkennung von Webservern gibt. Um die vom Verfasser entwickelte Fingerprinting-Software für den Einsatz in HTTP zu benutzen, müssen die FTP-Kommandos durch ihre HTTP-Äquivalente ersetzt werden. Die Reintegration der relevanten Testfälle in Hmap geschieht auf die gleiche Weise wie bei FTPmap. Jedoch konnte der Verfasser während der Erstellung dieser Diplomarbeit keine Testfälle finden, die die Genauigkeit des HTTP-Fingerprintings verbesserten.

## *5.3 Bewertung der Zielerreichung*

Nachdem die Zuverlässigkeit der entwickelten Lösung mit quantitativen Methoden gezeigt wurde, sollen nun auch die qualitativen Anforderungen[78] bewertet werden.

**Portabilität**

Die verwendeten Programmiersprachen und Technologien sind für eine Vielzahl von Plattformen verfügbar und ermöglichen so eine leichte Portierung der entwickelten Software auf unterschiedliche Betriebssysteme.

**Erweiterbarkeit**

Fingerprints von Betriebssystemen und Diensten werden nicht hart kodiert sondern in einer dynamisch erweiterbaren Fingerprint-Collection gespeichert.

**Bedienbarkeit**

Die Anforderungen an die Bedienbarkeit der Software werden durch die Entwicklung einer übersichtlich gestalteten GUI auf Basis der Widgets-Bibliothek Tk erfüllt.

Wie dargelegt konnte nicht nur die Zuverlässigkeit von bestehenden Fingerprinting-Ansätzen verbessert werden, sondern auch die selbst definierten Anforderungen an die Software-Lösung erfüllt werden. Dies ermöglicht dem Verfasser, ein positives Fazit aus der Bearbeitung des Themas zu ziehen.

---

[78]   Vgl. Kapitel 3.5

# 6 Fazit und Ausblick

Die Machbarkeit von Fingerprinting mittels Fuzzing wurde theoretisch erörtert, begründet und durch eine konkrete Implementierung unter Beweis gestellt. In den durchgeführten Tests wurde eine zuverlässige Unterscheidung verschiedener Betriebssysteme und Dienste demonstriert. Dabei wurden in zwei Fällen reproduzierbare Unterschiede im Antwort-Verhalten beobachtet, die von bestehender Fingerprinting-Software noch nicht erfasst werden.

Durch die Anwendung von Fuzzing zur Erzeugung von Anfrage-Nachrichten können derartige Verhaltens-Unterschiede auch in zukünftigen Releases von Betriebssystemen und Diensten systematisch erforscht werden. Zur Verbesserung der bestehenden Fingerprinting-Software können die relevanten Anfrage-Nachrichten aus den gesammelten Fingerprints extrahiert und der eigenen Anfragenmenge hinzugefügt werden.

Gegenüber der Technik des Banner-Grabbings, die Versionsangabe eines Dienstes beim Verbindungsaufbau zu untersuchen, ist die vorgestellte Lösung im Vorteil, wenn diese Versionsangabe deaktiviert oder verfälscht wurde. Während Banner-Grabbing-Software in diesen Fällen zu keinen oder vom Anwender unbemerkt falschen Ergebnissen kommt, ist eine Unterscheidung durch die hier vorgestellte Fingerprinting-Lösung weiterhin möglich.

Die nächste Entwicklungsstufe des OS-Fingerprintings wäre die Analyse von Häufungen und Sequenzen im Antwort-Verhalten, die, wie im Konzept erläutert, eine weiter differenzierte Unterscheidung von Betriebssystemen ermöglicht.

Für die Zukunft des Service-Fingerprintings ist die Einbindung evolutionärer Fuzzing-Methoden denkbar. Durch Selektion und Rekombination besonders aussagekräftiger Testfälle könnte die Erfolgsquote des Fingerprintings entscheidend verbessert werden. Weitere Forschungsarbeit ist nötig, um einen geeigneten Ansatz für ein Fingerprinting auf Basis von Evolutionary Fuzzing zu finden.

# Anhang

## *Anhangverzeichnis*

## *Anhang 1: Spezifikation eines Pseudocodes*

Für die Darstellung der zugehörigen Algorithmen soll ein Pseudocode (angelehnt an Dalbey, John (2003)) verwendet werden, um die Abläufe unabhängig von einer konkreten Implementierung zu verdeutlichen. Dabei gelten folgende Konventionen:

- Eine Funktion wird durch Angabe ihres Bezeichners und der übergebenen Variablenliste in Klammern spezifiziert. Nach einem Doppelpunkt folgt der Rumpf der Funktion. Die Funktion endet mit einem return-Statement, durch das Werte zurückgegeben werden können.

  *procedure* ( *Variables List* ):
  > *Function Body*
  > **return** *value*

- Verzweigungen und Zählschleifen erfolgen über die hinlänglich bekannten Konstrukte if und for, deren Syntax im Folgenden gezeigt werden soll.

  **if** ( *Condition* )
  > *Statements*
  **endif**

  **for** *iterator* = *start* **to** *end*
  > *Statements*
  **end-for**

- Variablen werden durch := zugewiesen.
- Arrays (Felder) werden durch eckige Klammern indiziert.
- Eine Funktion wird durch ihren Bezeichner und die Liste übergebener Variablen (in Klammern) aufgerufen.

# Quellenverzeichnis

AktG, Aktiengesetz idF. vom 12. August 2008 (BGBl. I S. 1666, 1668)

Arkin, Ofir (2003), „Revolutionizing Active Operating System Fingerprinting - The Present & Future Xprobe2", Black Hat 2003, http://www.blackhat.com/presentations/bh-usa-03/bh-us-03-arkin.pdf, Stand: 10. Juli 2008

Arzur, Erwan (2005), „TCP Timestamp and Advanced Fingerprinting", http://www.se-curiteam.com/securityreviews/5LP0K2KF6I.html, Stand: 4. September 2008

Asteroth, Alexander; Baier, Christel (2003), „Theoretische Informatik", München 2003

Bellovin, Steven M. (1989), „Security Problems in the TCP/IP Protocol Suite", SIG-COMM Comput. Commun. Rev., ACM, 1989.

Bellovin, Steven M. (2002), „A Technique for Counting NATted Hosts", in Proceedings of the 2nd ACM SIGCOMM Workshop on Internet measurement, ACM, 2002.

BSI (2003), „Durchführungskonzept für Penetrationstests", http://www.bsi.bund.de/lite-rat/studien/pentest/penetrationstest.pdf, Stand: 24. Juni 2008

BSI (2007a), „Leitfaden IT-Sicherheit", http://www.bsi.bund.de/gshb/Leitfaden/GS-Leitfaden.pdf, Stand: 24. Juni 2008B.bundestag.de/dip21/brd/2007/0798-07.pdf, Stand: 25. Juni 2008

Bursztein, E. (2007), „Time has something to tell us about Network Address Translati-on", http://www.lsv.ens-cachan.fr/Publis/PAPERS/PDF/Bur-nordsec07.pdf, Stand: 9. September 2007

Caballero, J.; Venkataraman, S.; Poosankam, P.; Kang, M. G.; Song, D.; Blum, A. (2007), „Automatic fingerprint generation",  Network and Distributed System Security Symposium, 2007.

Dalbey, John (2003), „PSEUDOCODE STANDARD",
http://users.csc.calpoly.edu/~jdalbey/SWE/pdl_std.html, Stand: 29. Juli 2008

Engesser, Hermann; Claus, Volker; Schwill, Andreas (1993), „Duden Informatik", 2. Aufl., Mannheim 1993

Firewall.cx (2008), „TCP Options", http://www.firewall.cx/tcp-analysis-section-6.php, Stand: 10. September 2008

Gagnon, Francois; Esfandiari, Babak; Bertossi, L., (2007) „A Hybrid Approach to Operating System Discovery using Answer Set Programming", Integrated Network Management, IEEE, 2007.

Glaser, Thomas (2000), „TCP/IP Stack Fingerprinting Principles",
http://www.sans.org/resources/idfaq/tcp_fingerprinting.php, Stand: 4. Juli 2008

Greenwald, Lloyd G.; Thomas, Tavaris J. (2007),„Toward undetected operating system fingerprinting", in Proceedings of the first USENIX workshop on Offensive Technologies, USENIX, 2007.

GmbHG, GmbH-Gesetz idF. vom 19. April 2007 (BGB1. I S. 542, 547)

Haake, Jörg; Icking, Christian; Landgraf, Britta; Schümmer, Till (2007), „Verteilte Systeme", Universitätskurs, FernUniversität in Hagen, Hagen 2007

IANA (2008), „PORT NUMBERS", http://www.iana.org/assignments/port-numbers, Stand: 5. Juli 2008

IETF (1985), „FILE TRANSFER PROTOCOL (FTP)", http://tools.ietf.org/html/rfc959, Stand: 23. Juli 2008

Jakobs, Holger (2004), „Kleine Einführung in Tcl",
https://www.bg.bib.de/portale/bes/pdf/Einfuehrung_Tcl.pdf, Stand: 19. Juli 2008

Johansson, Jesper; Riley, Steve, (2005), „Anatomy Of A Hack", The Ethical Hacker
Network, 2005, http://www.ethicalhacker.net/content/view/8/2/, Stand: 15. Juni 2008

Kaiser, Peter; Ernesti, Johannes (2007), „Python - Das umfassende Handbuch", Bonn
2007

Lippmann, R.; Fried, D.; Piwowarski, K.; Streilein, W. (2003)
„Passive operating system identification from TCP/IP packet headers", in Proceedings
of the ICDM Workshop on Data Mining for Computer Security, 2003.

McKinney, David (2005), „Antiparser", http://antiparser.sourceforge.net/, Stand: 23.
Juli 2008

Miller, Barton P. (1989), „An Empirical Study of the Reliability of UNIX Utilities",
ftp://ftp.cs.wisc.edu/paradyn/technical_papers/fuzz.pdf, Stand: 13. Juni 2008

Miller, Charlie (2007), „Real World Fuzzing", ToorCon 2007,
http://toorcon.org/2007/talks/60/real_world_fuzzing.pdf, Stand: 22. Juli 2008

Mörike, Michael (Hrsg.) (2004), „IT-Sicherheit", Wolfsburg 2004

Poppa, Ryan (2007), „Where have all the good fingerprinters gone?", http://numeropho-
be.com/?p=34, Stand: 7. Juli 2008

Ruef, Marc (2007), „Die Kunst des Penetration Testing", Böblingen 2007

Rütten, Christiane (2007), „Datensalat", Heise Security,
http://www.heise.de/security/Schwachstellensuche-mit-Fuzzing--/artikel/100167/0,
Stand: 7. Juli 2008

Sanfilippo, Salvatore (2005), „Hping – Active Network Security Tool",

http://hping.org/, Stand: 19. Juli 2008

Sarraute, Carlos; Burroni, Javier (2008), „Using Neural Networks to improve classical Operating System Fingerprinting techniques", Electronic Journal of SADIO, Vol. 8, No. 1, 2008.

Singh , Pukhraj; Mookhey, K.K., SecurityFocus (2004), „Metasploit Framework, Part 1", http://www.securityfocus.com/infocus/1789, Stand: 22. Juli 2008

Smart, Matthew; Malan, G. Robert; Jahanian, Farnam (2000), „Defeating TCP/IP Stack Fingerprinting", in Proceedings of the 9th USENIX Security Symposium. Denver, Colorado, USA, USENIX 2000.

Speichert, Horst (2007), „Praxis des IT-Rechts", 2. Aufl., Wiesbaden 2007

Stevens, W. Richard (1994), „TCP/IP Illustrated", Volume 1, 18. Aufl., Upper Saddle River NJ 1994

Sutton, Michael; Greene, Adam; Amini, Pedram. (2007), „Fuzzing, Brute Force Vulnerability Discovery", Upper Saddle River NJ 2007

Tanenbaum, Andrew S. (1992), „Computer-Netzwerke", 2. Aufl., Attenkirchen 1992

Tanenbaum, Andrew S.; Steen, Maarten van (2007), „Distributed Systems", 2. Aufl., Upper Saddle River NJ 2007

Vaskovich, Fyodor (1999), „Das Erkennen von Betriebssystemen mittels TCP/IP Stack FingerPrinting", http://nmap.org/nmap-fingerprinting-article-de.html, Stand: 15. Juni 2008

Zalewski, Michael (2006), „Dr. Jekyll had something to Hyde", http://lcamtuf.coredump.cx/p0f/README, Stand: 18. Juli 2008